%% file: Article_SigmaStar_XiStar_pp7TeV.tex
\documentclass[ALICE,manyauthors]{cernphprep}
\usepackage{lmodern}
\usepackage{epstopdf}
\usepackage{lineno}

\usepackage{caption}
\usepackage{footnote}

\usepackage{graphics}
\usepackage{multirow}
\usepackage{wasysym}

\usepackage{hyperref}
\usepackage[numbers,sort&compress]{natbib}

\newcommand{\s}{$\sqrt{s}$}
\newcommand{\pT}{\ensuremath{p_{\rm T}}}
\newcommand{\dedx}{d$E$/d$x$}
\newcommand{\dndy}{d$N$/d$y$}
\newcommand{\dndydpt}{${\rm d}^2N/({\rm d}y {\rm d}p_{\rm T})$}
\newcommand{\pp}{pp}

\newcommand {\chisquare} {$\chi^{2}$}
\newcommand {\reducedchisquare} {$\chi^{2}/\mathrm {ndf}$}

\newcommand {\gmom}   {\mbox{\rm GeV$\kern-0.15em /\kern-0.12em c$}}
\newcommand {\mmom}   {\mbox{\rm MeV$\kern-0.15em /\kern-0.12em c$}}
\newcommand {\Gmass} {\mbox{\rm GeV$\kern-0.15em /\kern-0.12em c^2$}}
\newcommand {\mmass} {MeV/$c^{\mathrm 2}$}
\newcommand {\modrap} {$\left | y \right | $}

\newcommand {\meanpT}{$\langle \pT \rangle$}
\newcommand{\rmSigma}          {\mbox{$\mathrm {\Sigma}$}}
\newcommand{\rmASigma}          {\mbox{$\mathrm {\overline{\Sigma}}$}}
\newcommand{\rmLambda}          {\mbox{$\mathrm {\Lambda}$}}
\newcommand{\rmALambda}         {\mbox{$\mathrm {\overline{\Lambda}}$}}
\newcommand{\rmXiZres}  {\mbox{$\mathrm {\Xi (1530)^{0}}$}}
\newcommand{\rmAXiZres}  {\mbox{$\mathrm {\overline{\Xi} (1530)^{0}}$}}
\newcommand{\piPlus}            {\mbox{$\mathrm {\pi^+}$}}
\newcommand{\piMinus}            {\mbox{$\mathrm {\pi^-}$}}
\newcommand{\piPlusMinus}       {\mbox{$\mathrm {\pi^\pm}$}}

\newcommand{\rmXi}      {\mbox{$\mathrm {\Xi}$}}

\newcommand{\Tsallis}   {L\'{e}vy-Tsallis}

\renewcommand{\thefootnote}{\textdagger}

\begin{document}
%
%
\begin{titlepage}
\PHyear{2014}
\PHnumber{128}                 
\PHdate{06 June}              
%
%
\title{Production of $\mathbf{\Sigma(1385)^{\pm}}$ and $\mathbf{\Xi(1530)^{0}}$\\ in proton-proton collisions at $\mathbf{\sqrt{s}=}$ 7 TeV}
\ShortTitle{Production of $\Sigma(1385)^{\pm}$ and $\Xi(1530)^{0}$}   %
%
\Collaboration{ALICE Collaboration%
         \thanks{See Appendix~\ref{app:collab} for the list of collaboration
                      members}}
\ShortAuthor{ALICE Collaboration}      
\begin{abstract}
The production of the strange and double-strange baryon resonances ($\Sigma(1385)^{\pm}$, $\Xi(1530)^{0}$) 
has been measured at mid-rapidity (\modrap$<0.5$) in proton-proton collisions at \s~=~7~TeV with the ALICE detector at the LHC. 
Transverse momentum spectra for inelastic collisions 
are compared to \mbox{QCD-inspired} models, which in general underpredict the data.
A search for the $\phi(1860)$ pentaquark, decaying in the $\Xi\pi$ channel, has been carried out but
no evidence is seen.
\end{abstract}
\end{titlepage}
\setcounter{page}{2}
%
\section{Introduction}
\label{intro}
The study of strange baryon resonances in proton-proton (\pp) collisions 
contributes to the understanding of hadron production mechanisms and 
provides a reference for tuning QCD-inspired event generators.
The strange-quark content makes these baryons a valuable tool in understanding
production mechanisms, since the initial state
colliding projectiles contain no strange valence quarks and therefore
all strange particles are created in the collision. 

In addition, a measurement of resonance production in the
\pp~system serves as a reference for understanding resonance production 
in heavy-ion collisions, where resonances, due to their
lifetime of a few fm/$c$ being comparable to the lifetime of the hadronic phase,
are sensitive probes of the dynamical evolution of the fireball. 
Previous measurements at a collision energy of \s~=~0.2~TeV
with the STAR detector at the RHIC have shown 
that the yields of \rmSigma(1385) in Au--Au in comparison
to \pp~collisions indicate the presence of rescattering and regeneration 
in the time span between chemical and kinetic freezeout~\cite{STAR_StrangeBaryons}. 
Forthcoming analysis of strange baryon resonances in Pb--Pb collisions by the ALICE collaboration
will further explore those effects at higher energy and density of the colliding system.
The results for the \rmSigma(1385)$^{\pm}$ and \rmXiZres~baryons in \pp~collisions 
will therefore serve as benchmark.

Measurements of differential (\dndydpt) and integrated
(\dndy) yields of the \rmSigma(1385)$^{\pm}$
and \rmXiZres~baryons are presented at mid-rapidity (\modrap$<0.5$) in
inelastic (INEL) \pp~collisions at \s~=~7~TeV, collected with the ALICE detector~\cite{ALICE_JINST} 
at the LHC. 
The differential spectra are compared to Monte Carlo (MC) event generators.
The mean transverse momentum \meanpT~is compared to those of
other particles measured in \pp~collisions with the ALICE detector 
at both \s~=~7~TeV and \s~=~0.9~TeV, and with
the STAR detector at \s~=~0.2~TeV. 

The  $\Xi$(1530) reconstruction channel $\Xi\pi$ is additionally analysed to investigate evidence
of the $\phi(1860)$ pentaquark, previously reported by the NA49 experiment~\cite{NA49}. 
No such signal was observed by other experiments at different energies 
and with different beams and  \linebreak reactions~\cite{ALEPH,BABAR,CDF,COMPASS,E690,FOCUS,HERA-B,HERMES,WA89,ZEUS,H1}.

This article is organized as follows. Section~\ref{experiment} gives a 
brief description of the main detectors used for this analysis and 
the experimental conditions.
Section~\ref{selections} describes track and topological selections.
Signal extraction methods are presented in Section~\ref{signal}, and 
the efficiency corrections in Section~\ref{efficiencies}.
The evaluation of systematic uncertainties is discussed in Section~\ref{systematic}. 
In Section~\ref{results}, the \pT~spectra and the integrated
yields of the studied particle species are given and compared to model
predictions. In Section~\ref{pentaquark} the search for the $\phi(1860)$ pentaquark
is discussed. 
Conclusions are presented in Section~\ref{conclusions}.

\section{Experiment and data analysis}
\label{experiment}
The ALICE detector~\cite{ALICE_JINST} 
is designed to study a variety of colliding systems,
including \pp~and lead-lead (Pb--Pb) collisions,
at TeV-scale energies. The sub-detectors used in this analysis are
described in the following. A six-layer silicon Inner
Tracking System (ITS)~\cite{ALICE_all} and a large-volume Time Projection Chamber (TPC)~\cite{Alme:2010}
enable charged particle reconstruction with excellent momentum
and spatial resolution in full azimuth down to a \pT~of 100 \mmom~in the 
pseudorapidity range $|\eta|<0.9$.
The primary interaction vertex is determined with the TPC and ITS detectors
with a resolution of 200 $\mu$m for events with few tracks ($N_{\rm{ch}}\simeq 3$) and below
100 $\mu$m for events with higher multiplicity ($N_{\rm{ch}}\gtrsim 25$).
In addition, both detectors are able to provide particle identification (PID) via energy-loss measurements.
The data analysis is carried out using a sample of  $\sim$~250 
million minimum-bias pp collisions at \s~=~7~TeV collected during 2010.
During the data-taking period, the luminosity at the interaction point was
kept in the range 
$0.6-1.2\times 10^{\rm 29}$~cm$^{\rm {-2}}$s$^{\rm {-1}}$.
Runs with a mean pile-up probability per event larger than 2.9\% are excluded from the analysis.
The vertex of each collision is required to be
within $\pm$10 cm of the detector's centre along the 
beam direction.
The event vertex range is selected to optimize the reconstruction 
efficiency of particle tracks within the ITS and TPC acceptance.

\subsection{Particle selections}
\label{selections}
The resonances are reconstructed via their hadronic decay channel,
shown in Table~\ref{tab:PDG} together with the branching ratio (BR). 
For \rmSigma(1385), all four charged species (\rmSigma(1385)$^{+}$, 
\rmSigma(1385)$^{-}$, \rmASigma(1385)$^{-}$ and \rmASigma(1385)$^{+}$) are
measured separately. 
\rmXiZres~is measured together with its antiparticle (\rmAXiZres)
due to limited statistics. Therefore in this paper, unless otherwise
specified, \linebreak \mbox{\rmXiZres~$\equiv ($\rmXiZres~$+$~\rmAXiZres$)/2$.}
Note that, for brevity, antiparticles are not listed and the selection criteria, described
in the following, are discussed for particles; equivalent criteria hold for antiparticles.
\begin{table}[h!]
\centering
\resizebox{\textwidth}{!}{   
\begin{tabular}{l|l|l|l|l|l}
\hline\noalign{\smallskip}
 & Valence quarks & Mass (\mmass) & Width/\it{c}$\tau$ & Decay channel & Branching ratio  (\%) \\
\hline\noalign{\smallskip}
\rmSigma(1385)$^{+}$ & uus & 1382.80 $\pm$ 0.35   & (36.0 $\pm$  0.7) \mmass& \rmLambda$+$\piPlus & 87.0 $\pm$ 1.5  \\
\rmSigma(1385)$^{-}$ & dds &   1387.2 $\pm$ 0.5   & (39.4 $\pm$  2.1) \mmass& \rmLambda$+$\piMinus &  87.0 $\pm$ 1.5 \\
\hline\noalign{\smallskip}
 \rmXiZres &  uss & 1531.80 $\pm$ 0.32 & (9.1 $\pm$ 0.5)  \mmass&  \rmXi$^{-}+$\piPlus & 66.7 \\ 
\hline\noalign{\smallskip}
\rmXi$^{-}$ & dss &  1321.71 $\pm$ 0.07 & 4.91 cm& \rmLambda$+$\piMinus & 99.887 $\pm$ 0.035 \\
\hline\noalign{\smallskip}
\rmLambda  & uds & 1115.683 $\pm$ 0.006 & 7.89 cm & p$+$\piMinus & 63.9 $\pm$ 0.5 \\
\hline\noalign{\smallskip}
\noalign{\smallskip}
\end{tabular}
} 
\caption{Particles involved in this analysis and their PDG parameters~\cite{PDG2012}. Antiparticles are not listed for brevity.
From~\cite{PDG2012}, \rmXiZres$\longrightarrow\Xi + \pi$ has a branching ratio of $\sim$~100\%, then 
\rmXiZres$\longrightarrow$\rmXi$^{-}+$\piPlus~has a branching ratio of $\sim$~66.7\% due to isospin considerations.}
\label{tab:PDG}    
\end{table}

Several quality criteria, summarized in Table~\ref{tab:selections}, are used for track selection.
%
\begin{table}[b!]
\centering
\begin{tabular}{ll}
\hline\noalign{\smallskip}
Common selections & \\
\hline\noalign{\smallskip}
$|\eta|$ & $<0.8$ \\
\pT & $> 0.15$~\gmom \\
number of TPC clusters & $>70$ \\
$\chi^{2}$ per cluster & $<4$ \\
\hline\noalign{\smallskip}
Primary track selections & \\
\hline\noalign{\smallskip}
DCA$_{z}$ to PV               & $<2$ cm \\
DCA$_{r}$ to PV               & $<7$~$\sigma_{\mathrm {DCA}}$(\pT) \\
number of SPD clusters & $\ge1$ \\
\hline\noalign{\smallskip}
PID (\rmSigma(1385) analysis only) & \\
\hline\noalign{\smallskip}
$|$(d$E/$d$x)_{\rm{measured}}-$(d$E/$d$x)_{\rm{expected}}|$ & $<3$~$\sigma_{\rm{TPC}}$ \\
\hline\noalign{\smallskip}
\noalign{\smallskip}
\end{tabular}
\caption{Track selection criteria.
PV stands for ``primary vertex''. DCA${_r}$ and DCA$_{z}$ are the distances of
closest approach in the transverse plane and in the longitudinal direction, respectively.
}
\label{tab:selections}    
\end{table}
Charged pions from the strong decay of both \rmSigma(1385) and \rmXiZres~are 
not distinguishable from primary particles and therefore primary track selections
are used. They are requested to have a distance of closest approach (DCA) to the primary
interaction vertex of less than 2 cm along the beam direction and 
a DCA in the transverse plane smaller than 
7~$\sigma_{\mathrm {DCA}}$(\pT), where 
$\sigma_{\mathrm {DCA}}$(\pT)~=~(0.0026~+~0.0050~\gmom~$\times\;$\pT$^{-1}$)~cm is 
the parametrization which accounts for the \pT-dependent resolution of the DCA
in the transverse plane~\cite{ALICE_track_impact_parameter}.
Primary tracks are also required to have at least one hit in one of the two innermost 
layers of the ITS (silicon pixel detector, SPD)
and at least 70 reconstructed clusters in the TPC out of the maximum 159 available, 
which keeps the contamination from secondary and fake tracks small, while
ensuring a high efficiency and  good \dedx~resolution. 
Tracks close to the TPC edge or with transverse momentum
\pT~$< 0.15$~\gmom~are rejected because the resolution of track reconstruction deteriorates.

In the \rmSigma(1385) analysis, PID is implemented
for \piPlusMinus~and p from \rmLambda.
Particles are identified 
based on a comparison
of the energy deposited in the TPC drift gas and an expected value
computed using a Bethe-Bloch parametrization~\cite{Aamodt:2011}.
The filter is set to 3~$\sigma_{\rm TPC}$, 
where $\sigma$ is the resolution estimated by averaging over reconstructed tracks. 
An averaged value of $\sigma_{\rm TPC}$~=~6.5\% is found over all reconstructed tracks~\cite{Abelev:2012}. 
PID selection criteria are not applied in the \rmXi(1530) analysis as the combinatorial background 
is sufficiently removed through topological selection.

$\Lambda$ produced in the decay of \rmSigma(1385)
decays weakly into $\pi^{-}$p with $\it{c}\tau=$~7.89 cm~\cite{PDG2012}. 
These pions and protons do not originate from the primary collision vertex,
and thus they are selected using a DCA to the interaction point greater 
than 0.05 cm. 
At least 70 reconstructed clusters in the TPC are requested for these tracks.
Further selection criteria to identify \rmLambda~are applied on the basis of the
decay topology as described in~\cite{Aamodt:2011}. 
Selection criteria for \rmLambda~used in the \rmSigma(1385) analysis are 
summarized in Table~\ref{tab:selections_SigmaStar}.

$\Xi^{-}$ produced in the decay of the \rmXiZres~decays weakly
into $\Lambda\pi^{-}$ with $\it{c}\tau=$~4.91 cm~\cite{PDG2012}. 
Pions are selected from  
tracks with a DCA to the interaction point greater than 0.05 cm.
Pions and protons from $\Lambda$ are 
required to have a DCA  to the interaction point greater than 0.04 cm.
All pions and protons are requested to have 
at least 70 reconstructed clusters in the TPC.
Decay topologies for $\Xi^{-}$ and $\Lambda$ are 
used as described in~\cite{Aamodt:2011}. Selection criteria are 
summarized in Table~\ref{tab:selections_XiStar}.
\begin{table}[tbh!]
 \begin{minipage}{\linewidth}
\centering
\begin{tabular}{ll}
\noalign{\vskip 3mm}    
\hline\noalign{\smallskip}
$|y_{\rm{\Sigma}^{*}}|$ & $<0.5$ \\
DCA of \rmLambda~decay products to PV   &  $>0.05$ cm                 \\
DCA between \rmLambda~decay products    &  $<1.6$ standard deviations \\
DCA of \rmLambda~to PV             &  $<0.3$ cm                  \\
\rmLambda~cosine of pointing angle & $>0.99$                      \\
\rmLambda~fiducial volume (R$_{r}$)          & $1.4<$~R$_{r}<$~100~cm         \\
\rmLambda~invariant mass window    & $m_{\rm{PDG}}$~$\pm$~10 \mmass              \\
\hline\noalign{\smallskip}
\noalign{\smallskip}
\end{tabular}
\caption{Selection criteria 
used in the \rmSigma(1385) analysis.
PV stands for ``primary vertex''. 
R$_{r}$ is the transverse radius 
of the decay vertex.
}
\label{tab:selections_SigmaStar}    
\end{minipage}
   \begin{minipage}{\linewidth}
\centering
\begin{tabular}{ll}
\noalign{\vskip 5mm}    
\hline\noalign{\smallskip}
$|y_{\rm{\Xi}^{*}}|$ & $<0.5$ \\
DCA of \rmLambda~decay products to PV & $>0.04$ cm \\
DCA between \rmLambda~decay products  & $<1.6$ standard deviations\\
DCA of \rmLambda~to PV                 & $>0.07$  cm \\
\rmLambda~cosine of pointing angle     & $>0.97$    \\
\rmLambda~fiducial volume (R$_{r}$)              & $0.8<$~R$_{r}<$~100~cm         \\
$\Lambda$ invariant mass window        & $m_{\rm{PDG}}$~$\pm$~6~\mmass     \\
DCA of pion (from $\Xi^{-}$) to PV     & $>0.05$~cm \\
DCA between $\Xi^{-}$~decay products  & $<1.6$ standard deviations\\
$\Xi^{-}$~cosine of pointing angle     & $>0.97$    \\
$\Xi^{-}$~fiducial volume (R$_{r}$)              & $0.8<$~R$_{r}<$~100~cm         \\
$\Xi^{-}$ invariant mass window        &  $m_{\rm{PDG}}$~$\pm$~6~\mmass     \\
\hline\noalign{\smallskip}
\noalign{\smallskip}
\end{tabular}
\caption{
Same as Table~\ref{tab:selections_SigmaStar} but for the \rmXi(1530) analysis.
}
\label{tab:selections_XiStar}    
\end{minipage}
\end{table}
\\
All these criteria are optimized to obtain maximum signal
significance. Values for the significance are presented
in Section~\ref{countingsignal}.
%
\subsection{Signal extraction}
\label{signal}
\subsubsection{Combinatorial background and event-mixing}
\label{eventmixing}
\begin{figure}[h!]
 \begin{minipage}[c]{\linewidth}
  \subfigure{
    \resizebox{0.5\textwidth}{!}{
      \includegraphics{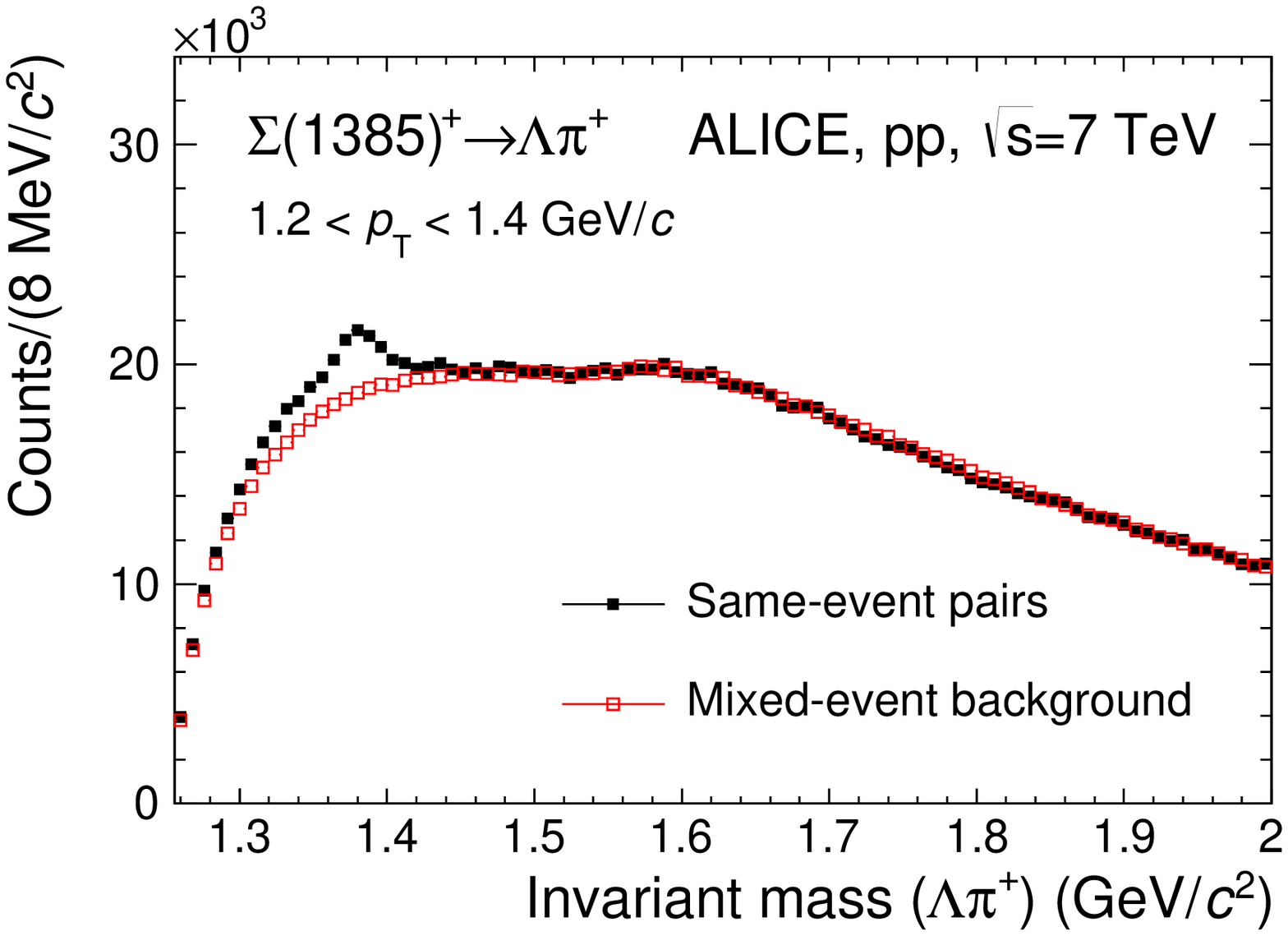}
    }
  } 
  \subfigure {
    \resizebox{0.5\textwidth}{!}{
      \includegraphics{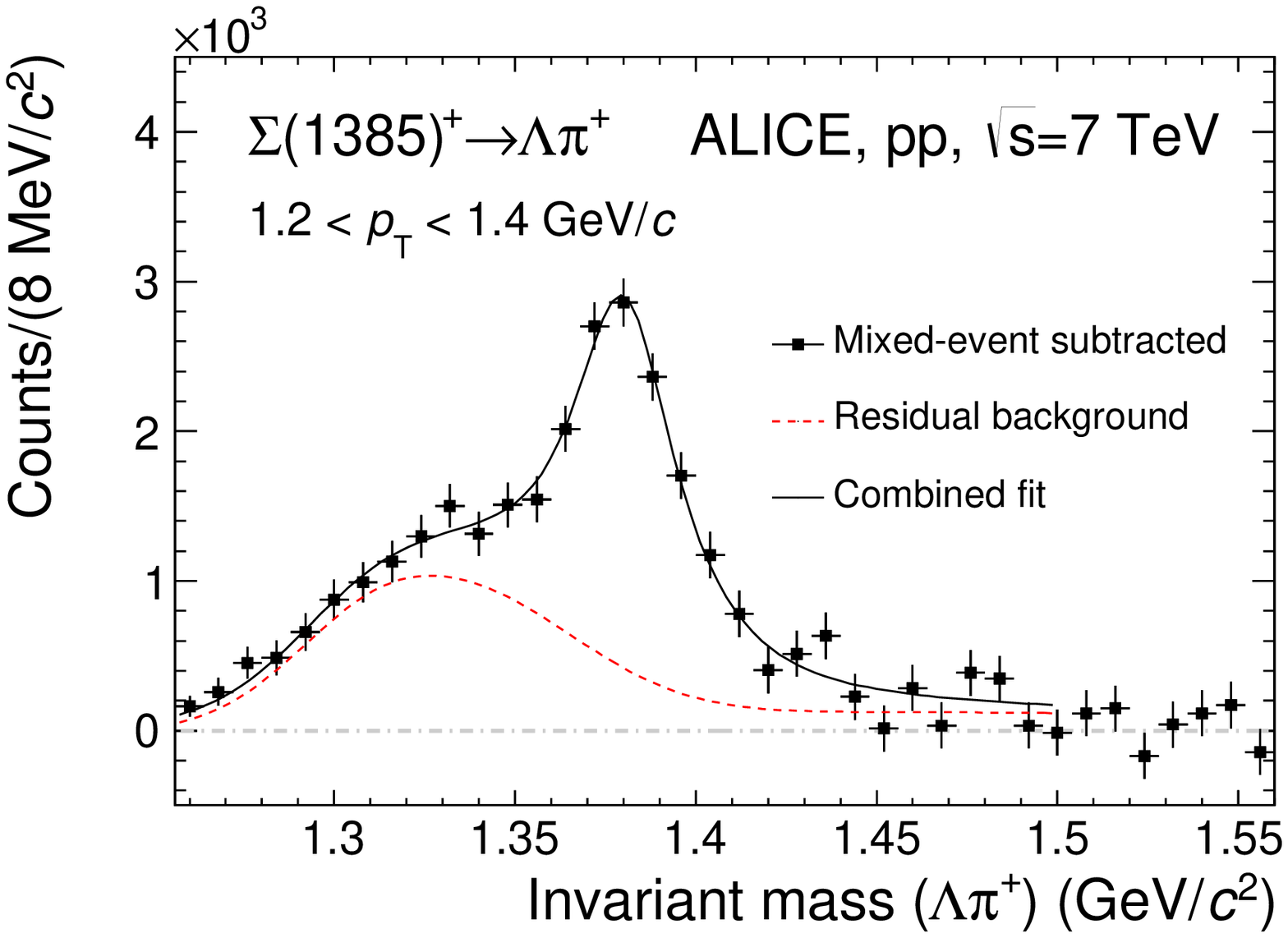}
    }
  }   
  \caption{(Colour online) 
    (Left panel) The \rmLambda\piPlus~invariant mass distribution in \modrap~$<$~0.5 for the transverse momentum bin 
    1.2~$<$~\pT~$<$~1.4~\gmom~ 
    in \pp~collisions at \s~=~7~TeV.
    The background shape estimated using pairs from different events (event-mixing) 
    is shown as open red squares. The mixed-event background is normalized
    in the range 1.56~$<M<$~2.0~\Gmass,
    where $M$ is the \rmLambda\piPlus~invariant mass.
    (Right panel) The  invariant mass distribution after mixed-event background subtraction 
    for 1.2~$<$~\pT~$<~$1.4~\gmom.
    The solid curve is the result of the combined fit (see text for details) and
    the dashed lines describes the residual background.
}
  \label{fig:invmassSigmaStarPlus}
%
%
  \subfigure{
    \resizebox{0.5\textwidth}{!}{
      \includegraphics{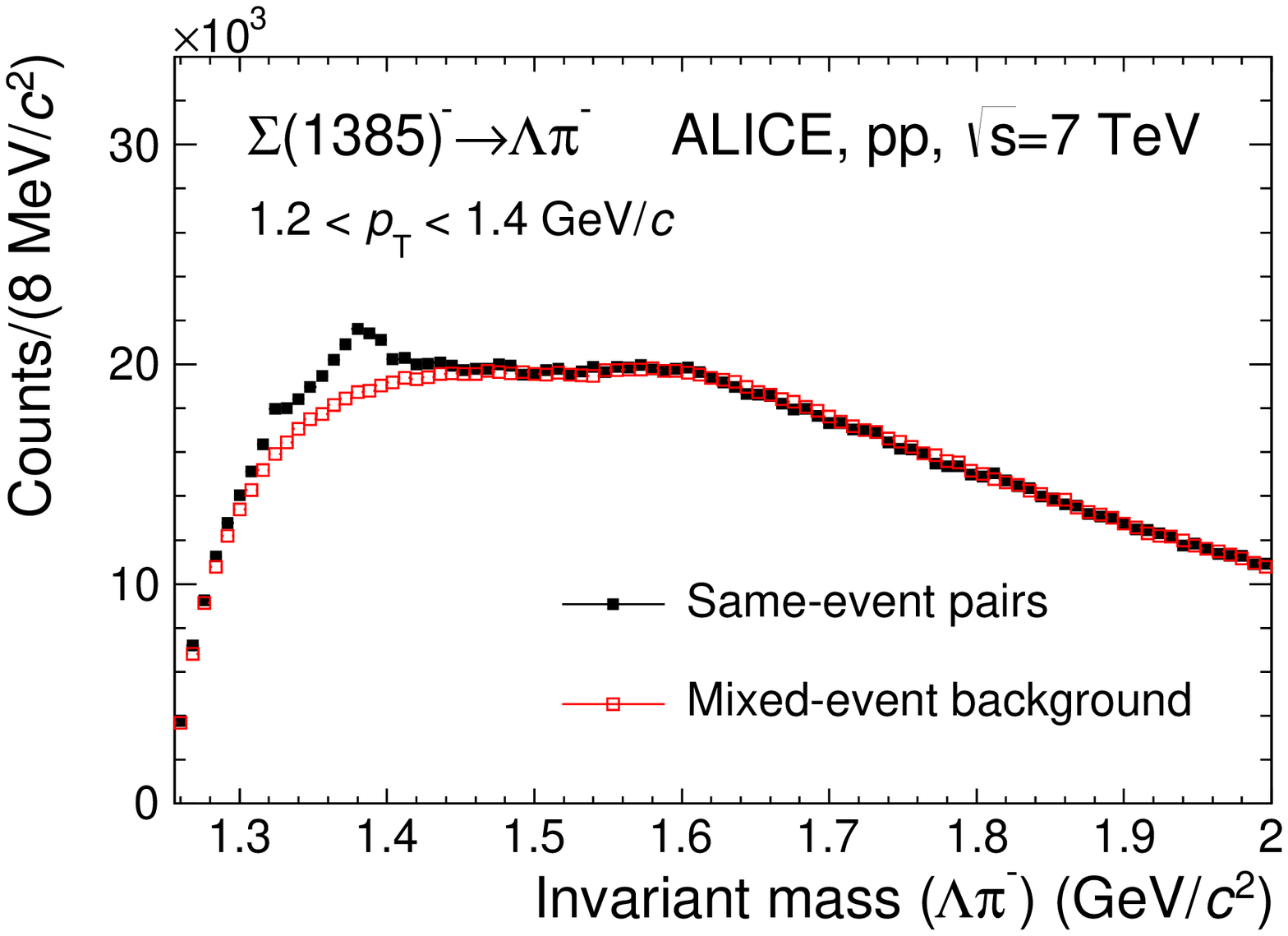}
    }
  } 
  \subfigure {
    \resizebox{0.5\textwidth}{!}{
      \includegraphics{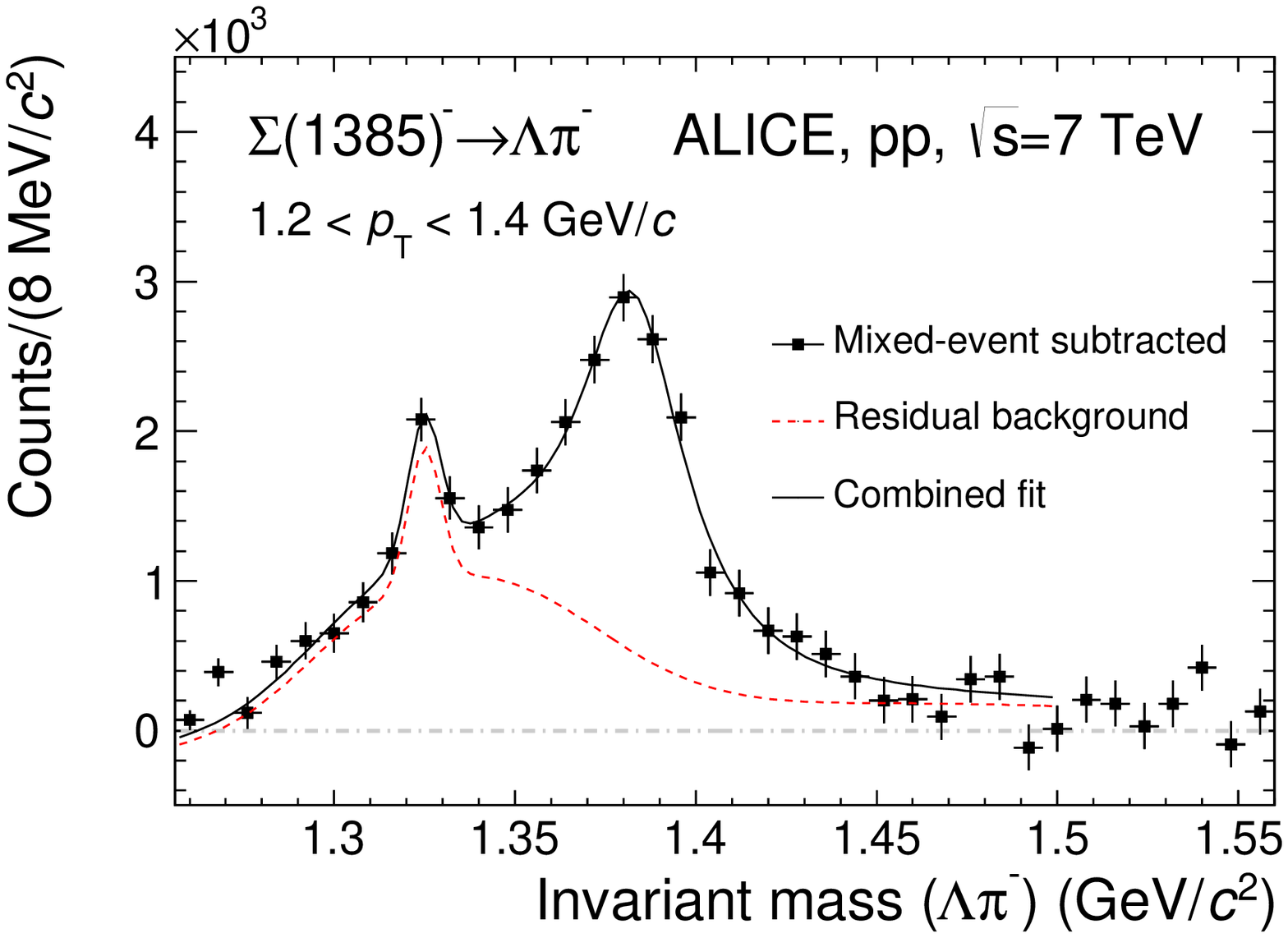}
    }
  }
 \caption{(Colour online) Same as Fig.~\ref{fig:invmassSigmaStarPlus} but for \rmSigma(1385)$^{-}$$\longrightarrow$\rmLambda$+$\piMinus. 
    Note the peak at around the $\Xi(1321)^{-}$ mass, which is absent in Fig.~\ref{fig:invmassSigmaStarPlus}.
  }
  \label{fig:invmassSigmaStarMinus}
\end{minipage}
\end{figure}

\begin{figure}[t]
  \subfigure {
    \resizebox{0.5\textwidth}{!}{
      \includegraphics{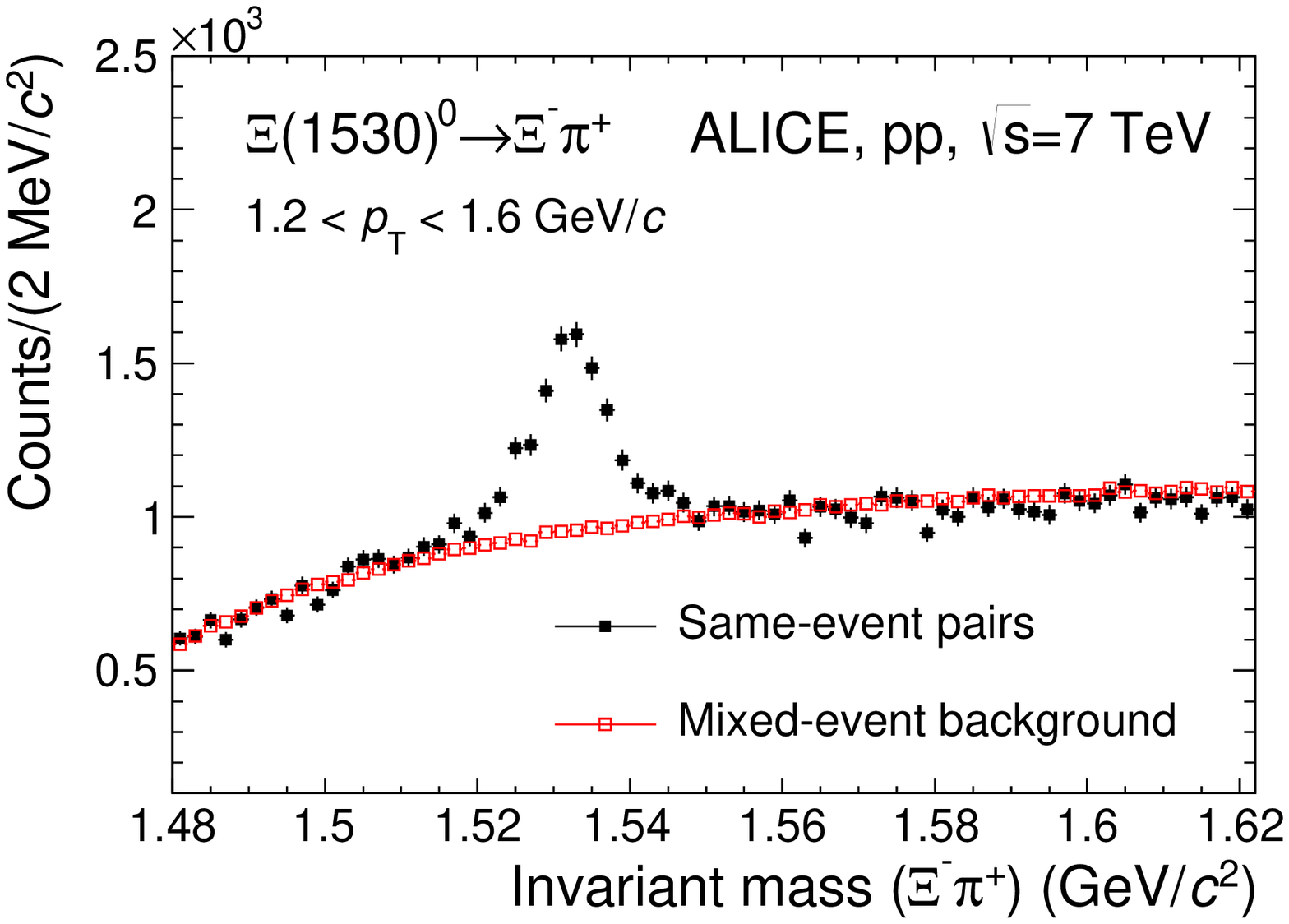}
  }
  } 
  \subfigure {
    \resizebox{0.5\textwidth}{!}{
      \includegraphics{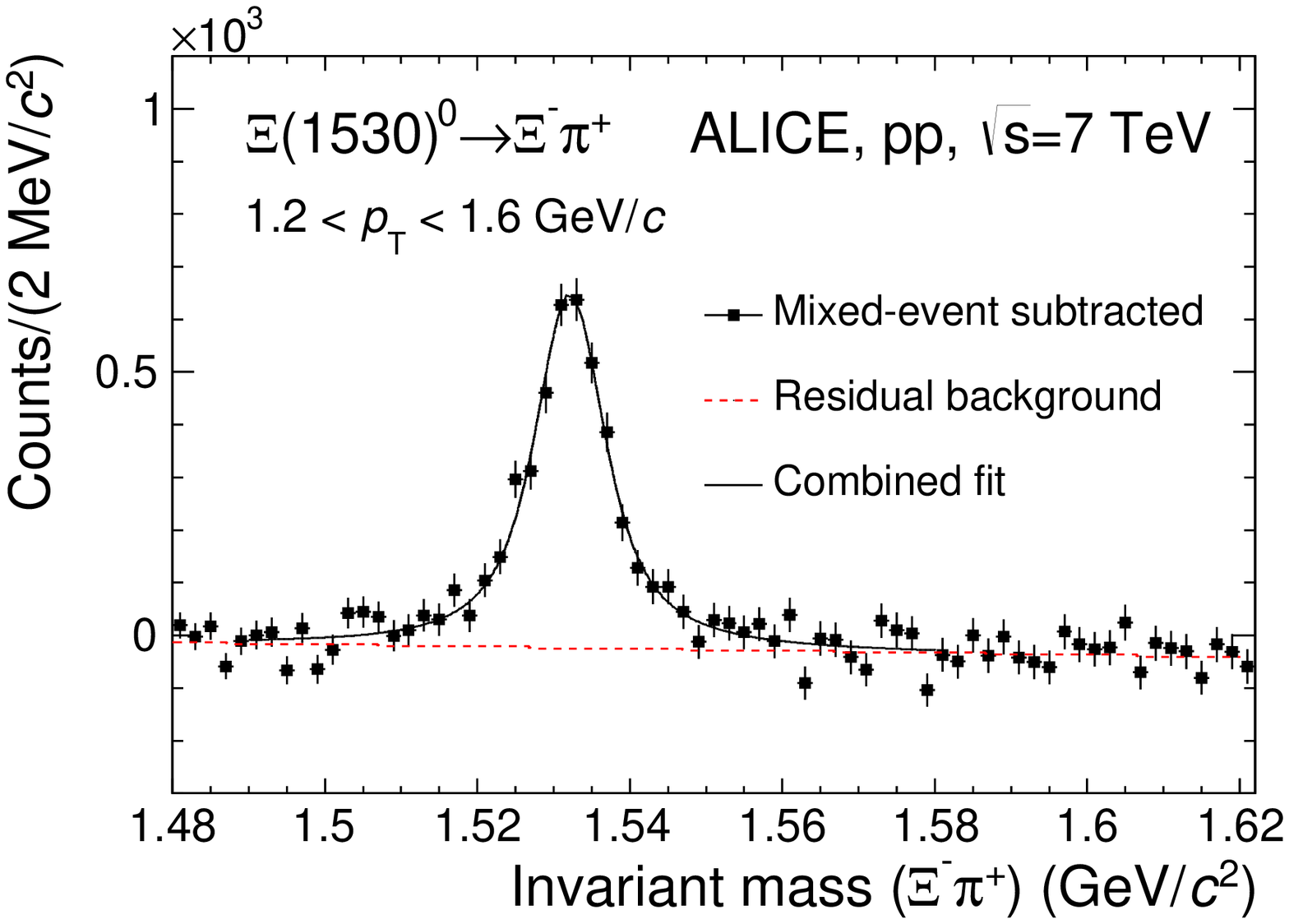}
    }
  } 
  \caption{(Colour online) 
    (Left panel) The \rmXi$^{-}$\piPlus~invariant mass distribution in \modrap~$<$~0.5 for the transverse momentum bin $1.2<$
    \pT$~<1.6$~\gmom~
    in \pp~collisions at \s~=~7~TeV.
    The background shape estimated using pairs from different events (event-mixing) 
    is shown as open red squares.  The mixed-event background is normalized
    in the range 1.49~$<M<$~1.51~\Gmass.
    (Right panel) The  invariant mass distribution after mixed-event background subtraction 
    for $1.2<$~\pT~$ <1.6$~\gmom.    
    The solid curve is the result of the combined fit and 
    the dashed line describes the residual background.}
  \label{fig:invmassXiStar}
\end{figure}

Due to their very short lifetime of a few fm/$c$, resonance decay products 
originate from a position that is indistinguishable from the primary vertex. 
Thus, the computation of invariant mass distributions for potential resonance
decay candidates has significant combinatorial background that has to be subtracted 
to ensure reliable yield determination.
This is shown in the left panels
of Figs.~\ref{fig:invmassSigmaStarPlus} and~\ref{fig:invmassSigmaStarMinus} 
(for \rmSigma(1385)$^{+}$ and \rmSigma(1385)$^{-}$, respectively) 
and Fig.~\ref{fig:invmassXiStar} (for the \rmXiZres).
Figures similar to Figs.~\ref{fig:invmassSigmaStarPlus} and~\ref{fig:invmassSigmaStarMinus} 
are obtained for the antiparticles \rmASigma(1385)$^{-}$ and \rmASigma(1385)$^{+}$.
In Fig.~\ref{fig:invmassSigmaStarMinus} the peak from $\Xi^{-}\longrightarrow\Lambda + \pi^{-}$ is visible.

The combinatorial background distributions are obtained and subtracted 
from the invariant mass distribution by means of a mixed-event technique, 
in which a reference background distribution is built with uncorrelated candidates from different events.
To avoid  mismatch due to different acceptances and to ensure a similar event 
structure, only tracks from events with similar vertex positions $z$ ($\Delta z <1$~cm) and track multiplicities $n$ 
($\Delta n <10$) are mixed. 
In order to reduce statistical uncertainties, each event is mixed with several other events
(5 in the \rmSigma(1385) analysis and $>20$ in the \rmXiZres~analysis), 
so that the total number of entries in the mixed-event invariant mass distribution is 
higher than the total number of entries in the distribution from the same event. 
Thus the mixed-event distribution needs to be scaled before it can be used to describe
the background in the same-event distribution.
For \rmSigma(1385), the regions for the normalization of the
mixed-event distribution are selected in the rightmost part of 
the invariant mass window, where the residual background is absent (see 
Section ~\ref{residual} for a description of the residual background). These regions
are different for the different \pT~bins, ranging from 1.48~$<M<$~2.0~\Gmass, 
for the lowest \pT~bin, to 1.95~$<M<$~2.0~\Gmass, for the highest \pT~bin 
($M$ being the invariant mass of \rmSigma(1385) and 2.0~\Gmass~being the upper extreme of the invariant mass
window). The reason for this \pT-dependent choice is due to the reach of the residual background, 
which is higher in invariant mass for higher \pT. 
Fixed regions, 1.6~$<M<$~1.8~\Gmass~and 1.8~$<M<$~2.0~\Gmass, have also been tried, giving a 
systematic uncertainty of $\sim$~1\%.
For \rmXiZres~a fixed region 1.49~$<M<$~1.51~\Gmass, just at the left of the signal, is selected.
A fixed region can be selected because for all \pT~intervals 
the background shape is similar and the invariant mass resolution 
on the reconstructed peak is the same. 
The uncertainty in the normalization ($\sim$~1\%), which is 
included in the quoted systematic uncertainty for signal extraction, 
is estimated by using another normalization region, 1.56~$<M<$~1.58~\Gmass, 
just at the right of the signal.
The open squares in the left panels
of Figs.~\ref{fig:invmassSigmaStarPlus},~\ref{fig:invmassSigmaStarMinus} and~\ref{fig:invmassXiStar} correspond to
the properly scaled mixed-event invariant mass distribution.
The right panels 
show the signals for each resonance after the mixed-event combinatorial background
is subtracted. 
\subsubsection{Residual correlated background}
\label{residual}
The mixed-event technique removes only uncorrelated background pairs in the 
invariant mass spectrum. The consequence is that residual correlations near
the signal mass range are not subtracted by the mixed-event spectrum and
correlated background pairs remain~\cite{HADES}.
This is especially dominant for \rmSigma(1385) 
(see Figs.~\ref{fig:invmassSigmaStarPlus} and~\ref{fig:invmassSigmaStarMinus}, right),
for which the correlated residual background takes contributions from 
two dominant sources:
\begin{itemize}
\item Type A: correlated $\Lambda\pi$ pairs coming from the decays of other particles which
have \rmLambda~and $\pi$ among the decay products.
\item Type B: correlated $\Lambda\pi$ pairs
which come from the dynamics of the collision and are not removed from the subtraction of 
the mixed-event background.  
\end{itemize}
All these contributions are present in the MC, albeit with potentially incorrect proportions. Thus,
simulations are used to determine the shapes of such contributions in invariant mass space and
then these contributions are renormalized using data, as described later.

All the sources of contamination of Type A, which can potentially produce correlated $\Lambda\pi$ pairs, 
are listed in Table~\ref{tab:contamination}. 
A similar scheme, not discussed for sake of brevity, 
is valid for the antiparticles (e.g. the $\overline{\Xi}^{+}$$\longrightarrow$\rmALambda\piPlus~decay
channel affects the reconstruction of \rmASigma(1385)$^{+}$).
Only sources A1, A5 and A6 in Table~\ref{tab:contamination} give a significant contribution 
to the correlated residual background of Type A. 
This is discussed in the following.
\begin{table}[h!]
\centering
  \begin{tabular}{l|ll|cc}
\hline\noalign{\smallskip}
&Source & BR &  \multicolumn{2}{c}{Affects}    \\
&        &      &  \rmSigma(1385)$^{+}$ & \rmSigma(1385)$^{-}$ \\ 
\hline\noalign{\smallskip}
A1 &$\Xi^{-}$$\longrightarrow$\rmLambda$$\piMinus &  99.9\% & &   \CheckedBox \\
\hline\noalign{\smallskip}
A2 &$\Xi(1530)^{-}\longrightarrow\Xi^{-}\pi^{0}$ & 33.3\% & &\multirow{2}{*}{\checked}\\
&\hspace{2.5cm} \rotatebox[origin=Tr]{330}{$\longrightarrow$}$\Lambda\pi^{-}$ & 99.9\% & &  \\
\hline\noalign{\smallskip}
A3 &$\Xi(1530)^{-}\longrightarrow\Xi^{0}\pi^{-}$ & 66.7\% & &\multirow{2}{*}{\checked } \\
&\hspace{2.5cm} \rotatebox[origin=Tr]{330}{$\longrightarrow$}$\Lambda\pi^{0}$ & 99.5\% & & \\
\hline\noalign{\smallskip}
A4 &$\Xi(1530)^{0}\longrightarrow\Xi^{-}\pi^{+}$ & 66.7\% & \multirow{2}{*}{\checked}  & \multirow{2}{*}{\checked}\\
&\hspace{2.5cm} \rotatebox[origin=Tr]{330}{$\longrightarrow$}$\Lambda\pi^{-}$ & 99.9\% & & \\
\hline\noalign{\smallskip}
A5 &\rmSigma(1385)$^{\pm}\longrightarrow\Sigma^{0}\pi^{\pm}$ & 5.8\% &   \multirow{2}{*}{\CheckedBox} & \multirow{2}{*}{\CheckedBox} \\
&\hspace{2.5cm} \rotatebox[origin=Tr]{330}{$\longrightarrow$}$\Lambda\gamma$ & 100\% &  &  \\
\hline\noalign{\smallskip}
A6 &$\Lambda(1520)\longrightarrow\Lambda\pi^{\pm}\pi^{\mp}$ & 5\% &  \CheckedBox  & \CheckedBox \\
\hline\noalign{\smallskip}
  \end{tabular}
\caption{Potential sources of contamination in the reconstruction of \rmSigma(1385). 
Checkmarks show which species is potentially affected. Checkboxes further
indicate whether the source gives a significant contamination (see text).
A similar scheme, not shown for sake of brevity, is valid for the antiparticles.}
\label{tab:contamination}    
\end{table}
\renewcommand{\thefootnote}{\fnsymbol{footnote}}

Source A1 in Table~\ref{tab:contamination} is due to the primary $\Xi^{-}$ which decays weakly to $\Lambda\pi^{-}$, 
affecting the reconstruction of \rmSigma(1385)$^{-}$. 
Since the $\Xi^{-}$ hyperon is metastable, it shows up in the $\Lambda\pi^{-}$ invariant mass spectrum as 
a very narrow peak at around the $\Xi^{-}$ mass, $M_{\rm{\Xi^{-}}}=$~1321.71~\mmass~\cite{PDG2012}, just 
on the left tail of the \rmSigma(1385)$^{-}$ signal. The  $\Xi^{-}$ peak is clearly seen in Fig.~\ref{fig:invmassSigmaStarMinus}. 
This contribution, which is
expected to be important since the yield of $\Xi^{-}$ is comparable to the yield of \rmSigma(1385)$^{-}$, 
is in fact suppressed, by an order of magnitude, 
because the filter on the DCA to the primary vertex of both $\Lambda$ and $\pi$
filters out most of the \rmLambda$\pi$ pairs from $\Xi^{-}$. 
Indeed, the filter on the DCA to the primary vertex is optimized for 
the \rmSigma(1385) decay products, which are not distinguishable from primary particles (see Section~\ref{selections}),
whereas $\Lambda$ and $\pi$ from $\Xi^{-}$ come from a secondary vertex, 
centimetres away from the primary vertex.
Only a small percentage of the $\Xi^{-}$ yield survives the filter on the DCA. Source A1 is taken
into account by adding a Gaussian function, with the mean value fixed to the $\Xi^{-}$ mass and the width and
normalization left free, 
to the combined fit of the invariant mass spectrum in the reconstruction of \rmSigma(1385)$^{-}$.
The contamination from $\Xi^{-}$ 
reaches about 5-10\%
of the raw \rmSigma(1385)$^{-}$ signal and varies little with \pT.

Sources A2, A3 and A4 give a negligible contribution. 
Sources A2 and A3 are due to the hadronic decay channels of $\Xi(1530)^{-}$, with BR = 33.3\% and 
BR = 66.7\%, respectively\footnote[1]{BR $\sim$~100\% for $\Xi(1530)\rightarrow\Xi\pi$~\cite{PDG2012},
then BR = ($\frac{1}{3}\times$100)\% for $\Xi(1530)^{-}\rightarrow\Xi^{-}\pi^{0}$ and 
BR = ($\frac{2}{3}\times100$)\% for $\Xi(1530)^{-}\rightarrow\Xi^{0}\pi^{-}$ due to isospin considerations.},  
and, like A1, affect only the \rmSigma(1385)$^{-}$ reconstruction.  
Source A4 is due to $\Xi(1530)^{0}$ and potentially affects the reconstruction of both 
\rmSigma(1385)$^{+}$ and \rmSigma(1385)$^{-}$, since it involves two opposite-sign pions.
The same topological considerations hold for A2 as they do for A1, since it involves a $\Xi^{-}$. 
Indeed, this $\Xi^{-}$ comes from 
the strong decay of $\Xi(1530)^{-}$, therefore it is practically not distinguishable from the (primary) $\Xi^{-}$ in A1. 
Unlike contribution A1, a further suppression, of about an order of magnitude with respect to A1, 
comes from both the smaller yield
of $\Xi(1530)^{-}$ with respect to the primary $\Xi^{-}$, and the BR of the $\Xi(1530)^{-}\rightarrow\Xi^{-}\pi^{0}$
channel. This further suppression makes contribution A2 practically negligible. 
Similar conclusions hold for contributions A3 and A4.

Source A5 in Table~\ref{tab:contamination} is related to the second $\Sigma(1385)$ decay channel, 
 $\Sigma(1385)^{\pm}\rightarrow\Sigma^{0}\pi^{\pm}$ (BR = $5.8$\%\footnote[2]{BR~=~($11.7 \pm 1.5$)\%~\cite{PDG2012} 
for $\Sigma(1385)\rightarrow\Sigma\pi$, then BR~=~($\frac{1}{2}\times$11.7)\% for the charged-pion channel 
$\Sigma(1385)^{\pm}\rightarrow\Sigma^{0}\pi^{\pm}$ 
due to isospin considerations.}),
with $\Sigma^{0}\rightarrow\Lambda\gamma$ (BR~$\simeq$~100\%~\cite{PDG2012}). $\Lambda$ from $\Sigma^{0}$
is paired with $\pi^{\pm}$ from $\Sigma(1385)^{\pm}$. This gives a Gaussian-like peak at around
1.306~\Gmass, with a
width of $\sim$~0.059~\Gmass~(FWHM). This peak is used in the combined fit to the signal (see below) 
with a relative normalization with respect to the signal which accounts for the ratio ($=0.067$) 
between the BR ($=5.8$\%) for the
$\Sigma(1385)^{\pm}\rightarrow\Sigma^{0}\pi^{\pm}$ channel and the BR ($=87$\%) 
for the $\Sigma(1385)^{\pm}\rightarrow\Lambda\pi^{\pm}$ channel.

Source A6 in Table~\ref{tab:contamination} is due to the $\Lambda(1520)\rightarrow\Lambda\pi^{\pm}\pi^{\mp}$ channel
(BR~=~$5$\%\footnote[3]{BR~=~$(10 \pm 1)\%$~\cite{PDG2012} for $\Lambda(1520)\rightarrow\Lambda\pi\pi$,
then BR~=~($\frac{1}{2}\times$10)\% for the charged-pions channel $\Lambda(1520)\rightarrow\Lambda\pi^{\pm}\pi^{\mp}$ 
due to isospin considerations.}).
The positive (negative) pion, paired with $\Lambda$,
produces a Gaussian-like peak, which contaminates the invariant mass distribution of \rmSigma(1385)$^{+}$ 
(\rmSigma(1385)$^{-}$). This peak is centred at $\sim$~1.315~\Gmass~and has  
a width of $\sim$~0.076~\Gmass~(FWHM).
The peak is used in the combined fit to the signal. The normalization of the peak is kept free
in the fit since the $\Lambda$(1520) yield is not measured.
The contamination from $\Lambda$(1520) decreases with increasing \pT, ranging
from about 75\% of the raw \rmSigma(1385)$^{-}$ signal in the first \pT~interval, down to 0 
for \pT~$>$~4~\gmom.

A third-degree polynomial is used to fit the residual background of Type B in the MC. 
The fit to MC data is performed in the region from 1.26~\Gmass~(just left of the signal region) 
to the lower edge of the event-mixing normalization region. 
The fitting function is then normalized to the residual background in real data;
the normalization is done in the region from 1.46~\Gmass~(just right of the 
signal region) to the lower edge of the event-mixing normalization region, 
where other sources of contamination are absent.
The lower point of the normalization region is the same
for all \pT~intervals since the mean, the width and 
the invariant mass resolution on the reconstructed peak stay the same over
all the \pT~range considered.
Comparable results are obtained from using different event generators  
(PYTHIA 6.4, tune Perugia~0~\cite{pythiatunes}, and PHOJET~\cite{PHOJET}) 
and other degrees for the polynomial (second and fourth).
The differences of about 2\% are included in the systematic uncertainties.  

The invariant mass distribution is fitted with a combined fit function: 
a (non-relativistic) Breit-Wigner peak plus the functions that make up the residual background
(Figs.~\ref{fig:invmassSigmaStarPlus} and~\ref{fig:invmassSigmaStarMinus}, right). 
The Breit-Wigner width $\Gamma$ is kept fixed to the PDG value to improve the stability of the fit.

For \rmXiZres, the residual background after the mixed-event background subtraction
is fitted with a first-degree polynomial. The fitting procedure is done in three stages. First, the background
is fitted alone from 1.48 to 1.59~\Gmass~while excluding the $\Xi$(1530)$^0$ mass 
region from 1.51 to 1.56~\Gmass. Second, a combined fit for signal and background 
is performed over the full
range with the background polynomial fixed to the results from the first fit stage;
a Voigtian function - a convolution of Breit-Wigner and Gaussian functions - is used for the signal. 
The Gaussian part accounts for detector resolution. 
Third, a fit is redone over the full range again with all parameters free but set
initially to the values from the second stage.
\subsubsection{Counting signal and signal characteristics}
\label{countingsignal}
The above procedure is applied for 10 (8) \pT~bins for \rmSigma(1385) 
(\rmXiZres), from 0.7 to 6.0 (0.8 to 5.6)~\gmom. 
For \rmSigma(1385), the fit is repeated leaving the Breit-Wigner width $\Gamma$ free to move,
and, for each \pT~interval, the difference in the yield is included in the systematic uncertainties ($\sim$~4\% maximum contribution).
The widths of both \rmSigma(1385) and~\rmXiZres~are consistent with the PDG values for all \pT~intervals.
In the \rmSigma(1385)$^{-}$~analysis, a Gaussian function, centred at 1.321~\Gmass~and with a starting value for the width of 2~\mmass, is 
used to help the combined fit around the $\Xi$(1321)$^{-}$ peak (Fig.~\ref{fig:invmassSigmaStarMinus}). 
The value of 2~\mmass~is obtained from the analysis of $\Xi$(1321)$^{-}$~\cite{Aamodt:2011} and is related to the mass resolution. 
Since the \rmSigma(1385)~mass binning of 8~\mmass, which is optimised for the \chisquare~of the combined fit, is larger than the mass resolution, 
only a rough description of the $\Xi$(1321)$^{-}$ peak is possible.
For \rmXiZres, the standard deviation of the Gaussian component of the 
Voigtian peak is found to be $\sim$~2~\mmass, which is consistent with the detector resolution, as obtained
from the MC simulation.
At low \pT, the fitted mass values  for \rmSigma(1385) are found to be slightly lower
(by $\sim$~5~\mmass) than the PDG value, which is attributed to imperfections 
in the corrections for energy loss in the detector material. For \rmXiZres,
the reconstructed masses are found to be in agreement with the PDG value within the statistical uncertainties.
\\The raw yields $N^{\rm RAW}$ are obtained by integrating the Breit-Wigner function. 
As an alternative, $N^{\rm RAW}$ is calculated by integrating the invariant mass histogram 
after the subtraction of the event-mixing background and 
subtracting the integral of the residual background (bin-counting method).
The difference between the two methods of integration is lower than 2\% on average.
\\
Significance values (defined as $S/\sqrt{S+B}$, 
where $S$ is the signal and $B$ the background) for \rmSigma(1385)$^+$
(\rmXiZres) are found to be 16.6 (16.5) in the lowest
\pT~interval, and 20.9 (22.8) in the highest \pT~interval, and reached
24.2 (52.4) in the intermediate \pT~interval.
Significance values comparable to those of \rmSigma(1385)$^+$ are obtained for the other \rmSigma(1385) species. 
%
%
\subsection{Correction and normalization}
\label{efficiencies}
\begin{figure}[h!]
\resizebox{1\textwidth}{!}{
   \includegraphics{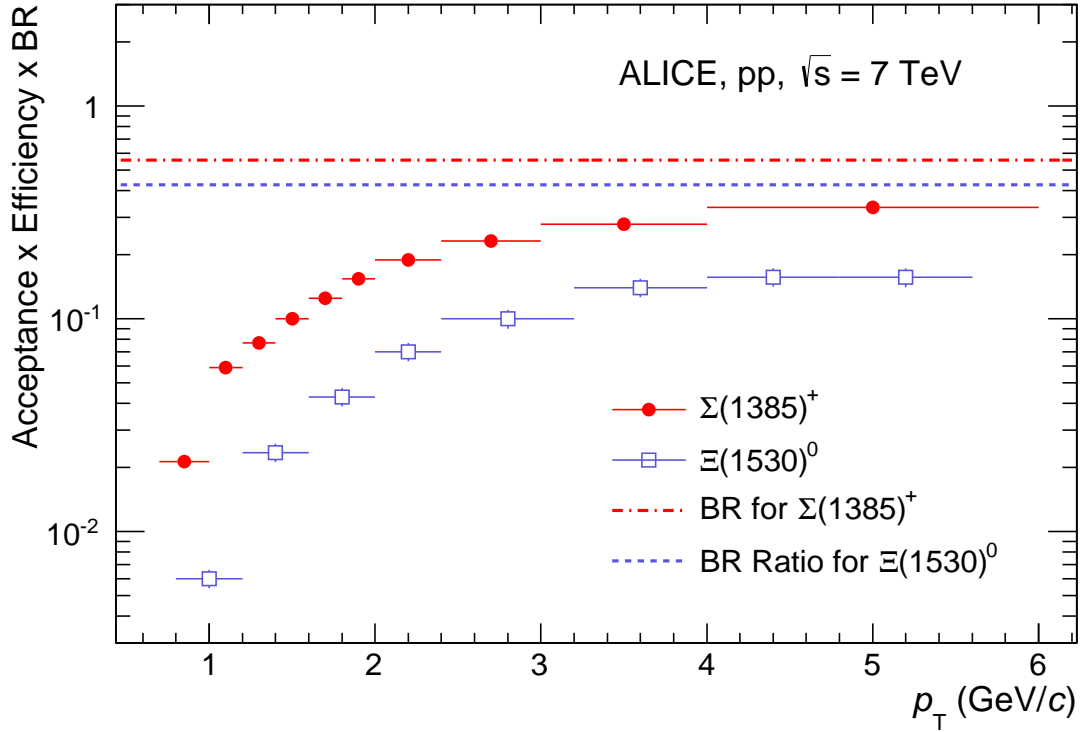}
}
\caption{ 
The product of acceptance, efficiency and branching ratio of \rmSigma(1385)$^+$ and \rmXi(1530)$^0$, 
obtained with PYTHIA 6.4~\cite{pythiatunes} and GEANT3~\cite{geant3},
as function of \pT~in \modrap$<$0.5.
Only statistical uncertainties are reported. 
The dashed- and the dash-dotted lines indicate the overall branching ratio for the two reconstruction channels.
}
\label{fig:eff}
\end{figure}

In order to extract the baryon yields, $N^{\rm RAW}$ are corrected for 
BR, the
geometrical acceptance ($A$), the detector efficiency ($\epsilon$)
and the correction factor which accounts for the GEANT3 overestimation of the $\bar{\rm p}$ 
cross sections ($\epsilon_{\rm GEANT3/FLUKA}$)~\cite{proton-antiproton}
\begin{equation} 
N^{\rm cor}(\pT)=\frac{N^{\rm RAW}(\pT)} {{\rm BR}\; (A\times\epsilon)(\pT)}\; \epsilon_{\rm GEANT3/FLUKA}(\pT).
\end{equation}
The product of acceptance and efficiency ($A \times \epsilon$) is determined 
from MC simulations with the PYTHIA 6.4 event generator (tune Perugia~0~\cite{pythiatunes}) 
and  a GEANT3-based simulation of the ALICE detector response~\cite{geant3}. 
The $\epsilon_{\rm GEANT3/FLUKA}$ correction factor is equal to 0.99 for the protons from 
\rmSigma(1385)$^{\pm}$ and \rmXiZres~and ranges from 0.90 to 0.98, from the lowest
to the highest \pT~interval, for the antiprotons from \rmASigma(1385)$^{\pm}$ and \rmAXiZres. 
About $200\times10^6$ Monte-Carlo events, with the same vertex distribution as for the real events,
were analysed in exactly the same way as for the data.
The $A\times\epsilon$  is determined from MC simulations as the ratio 
of the number of  reconstructed resonances 
to the number of those generated in \modrap$<0.5$, 
differentially as a function of transverse momentum,
as shown in Fig.~\ref{fig:eff}.
The drop in efficiency at low \pT~is due to the loss of slow pions involved in the decay chain. 
As a cross-check, the efficiency~$\times$~acceptance has also been assessed with PHOJET~\cite{PHOJET}
as event generator. The relative difference of the resulting $A \times \epsilon$,
averaged over the various \pT~intervals, is below 1\%.
\\Finally, corrections for the trigger inefficiency ($\epsilon_{\rm trigger}$)
and the loss of candidates outside of the $z$-vertex range ($\epsilon_{\rm vert}$) 
are applied via 
\begin{equation} 
\frac{1}{N_{\rm INEL}}\frac{\mathrm{d}^2N} {\mathrm{d}y\mathrm{d}\pT} = \frac{N^{\rm cor}(\pT)} 
{\Delta y \Delta \pT} \; \frac {\epsilon_{\rm trigger} } {\epsilon_{\rm vert} }\;
\frac{1} {N_ {\rm MB}},  
\end{equation}

\noindent where $N^{\mathrm {cor}}$ and  $N_ {\mathrm {MB}}$ are the number of 
reconstructed \rmSigma(1385) or \rmXi(1530) and the total number of minimum bias triggers, 
respectively. $\Delta$$y$ and $\Delta$\pT~are the rapidity window width and the \pT~bin width, respectively.
 The trigger selection efficiency for inelastic collisions
$\epsilon_{\mathrm {trigger}}$ is equal to $0.852 ^{+0.062} _{-0.030}$~\cite{INEL}. 
The loss of resonances due to the trigger selection, estimated by MC simulations, 
is negligible, less than 0.2\%.
The  $\epsilon_{\rm vert}$  correction factor accounts for resonance losses  
($\sim$~7\%) due to the requirement to have a primary vertex $z$ position in the range $\pm$10~cm.
%
\subsection{Systematic uncertainties of \pT~spectra}
\label{systematic}
Two types of systematic uncertainties in the 
particle spectra are considered: \pT-dependent 
systematic uncertainties, which are due to the 
selection efficiency and signal extraction
at a given \pT, and \pT-independent 
uncertainties due to the normalization to inelastic collisions
and other corrections.

The minimum and maximum values of the major contributions
to the point-to-point systematic uncertainties are listed in 
Table~\ref{tab:sys}.
\begin{table}[h!]
\centering
\begin{tabular}{lcc}
\hline\noalign{\smallskip}
Source of uncertainty & \rmSigma(1385)  & $\Xi$(1530) \\
\hline\noalign{\smallskip}
Point-to-point & &\\
\hline\noalign{\smallskip}
signal extraction & 8-11\% & 5-6\% \\
tracks selection & 7\% &  1-3\% \\
topological selection & 6-7\% & 3-4\% \\
PID efficiency & 4-6\% & - \\ 
\hline\noalign{\smallskip}
\pT-independent  && \\
\hline\noalign{\smallskip}
INEL normalization & $^{+7.3\%}_{-3.5\%}$ &  $^{+7.3\%}_{-3.5\%}$ \\
material budget & 4\% &  4\% \\
GEANT3/FLUKA correction & 2\% & 2\% \\
branching ratio & 1.5\%& - \\
\hline\noalign{\smallskip}
\end{tabular}
\caption{Summary of the systematic 
uncertainties in the \rmSigma(1385) and $\Xi(1530)$ differential yield, \dndydpt.}
\label{tab:sys}    
\end{table}
The uncertainties introduced by tracking, topology selection 
and PID are obtained by varying the selection criteria for the decay products.
To this purpose, the selection criteria listed in Tables~\ref{tab:selections},~\ref{tab:selections_SigmaStar}~and~\ref{tab:selections_XiStar}~are
changed by a certain amount which varies the raw yield in real data by $\pm$10\%.
The maximum difference between the default yield and the alternate
value obtained by varying the selection, is taken as systematic
uncertainty.
The uncertainties introduced by the signal extraction come from
several sources: normalization of the event-mixing background, 
fitting function and range of the residual background, 
signal fitting and integration. For \rmSigma(1385), the
contamination from the $\Lambda$(1520) introduced the largest
contribution ($\sim$~8\%). All the sources are
combined by summing in quadrature the uncertainties for
each \pT.

Among the \pT-independent uncertainties, 
the INEL normalization leads to a +7.3\% and -3.5\% uncertainty~\cite{INEL},
the determination of the material thickness traversed by the particles 
(material budget) introduces a 4\% uncertainty and
the use of FLUKA~\cite{fluka,fluka2} to correct the antiproton absorption cross 
section in GEANT3 leads to a further 2\% uncertainty~\cite{proton-antiproton}.
For \rmSigma(1385), a further 1.5\% comes from the uncertainty in the 
branching ratio. A summary of the \pT-independent uncertainties is presented
in Table~\ref{tab:sys}.
%
\section{Results}
\label{results}
\begin{figure}[t]
  \resizebox{1.\columnwidth}{!}{
    \includegraphics{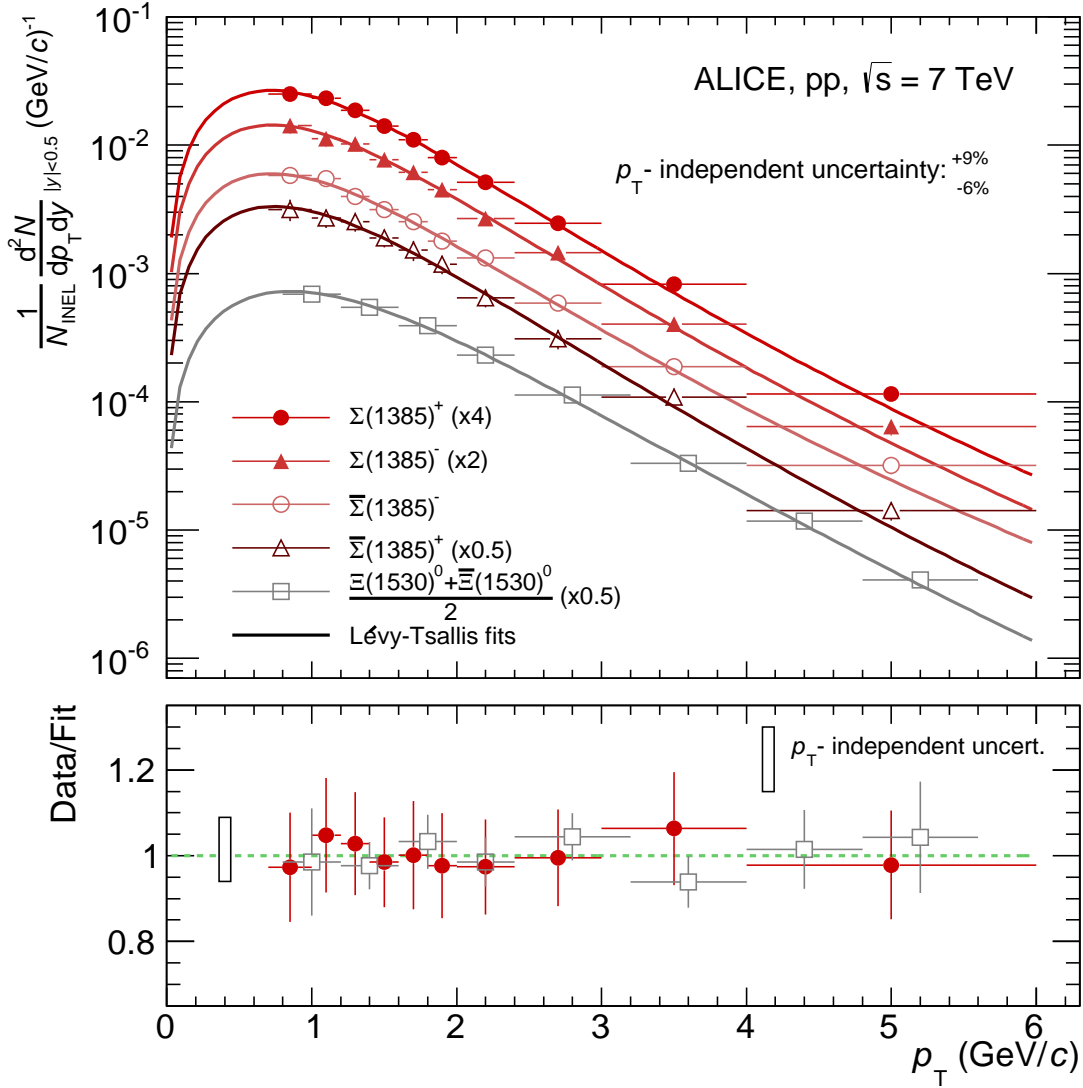}
  }
  \caption{Inelastic baryon yields, \dndydpt, per \pT~interval per unit rapidity for \rmSigma(1385) and
    \rmXiZres. Statistical
    and systematic uncertainties are summed in quadrature, excluding the \pT-independent uncertainties, 
    which affect only the overall normalization of the spectra and are not considered in the fit. 
    Spectra are fitted with a \Tsallis~function. The ratio data/fit is shown in the lower panel. 
    For the sake of visibility, only \rmSigma(1385)$^{+}$ is shown in the lower panel, but 
    similar ratios have been obtained for the other three \rmSigma(1385) species. For the ratio, 
    the integral of the fitting function in each corresponding \pT~interval is considered.
    Spectra points are represented at the centre of the \pT~interval.
  }
  \label{fig:ptspectrumSigmaStarAllSpecies}       
\end{figure}

The corrected baryon yields per \pT~interval per unit rapidity (1/$N_{\rm INEL}$ $\times$ \dndydpt) are shown in 
Fig.~\ref{fig:ptspectrumSigmaStarAllSpecies}. They cover the ranges 0.7 $<$ \pT~$<$~6.0~\gmom~for \rmSigma(1385) 
and 0.8 $<$ \pT~$<$~5.6~\gmom\linebreak for \rmXiZres. 
The vertical error bars in Fig.~\ref{fig:ptspectrumSigmaStarAllSpecies} represent the sum in quadrature
of the statistical and systematic uncertainties, excluding the \pT-independent uncertainties, 
which affect only the normalization.

All spectra are fitted with a \Tsallis~function~\cite{Tsallis:1988},
which is used for most of the identified particle spectra in pp collisions \cite{ALICE_charged, Abelev:2012, AliceMultiStrange, Spectra_900, Aamodt:2011},
\begin{eqnarray}
  \frac{1}{N_{\rm INEL}}\frac{\mathrm{d}^{2}N}{\mathrm{d}y\mathrm{d}\pT}  &  =  & \frac{(n-1)(n-2)} {nC [nC + m_{0}(n-2)]} \frac{\mathrm{d}N}{\mathrm{d}y} \;
  {\pT} \; \left( 1+ \frac{m_{\mathrm{T}}-m_{0}}{nC}\right)^{-n} \label{eqn:funclevy},
\end{eqnarray}
where $m_{\rm T}=\sqrt{m_{0}^2+\pT^2}$ and $m_{0}$ denotes the PDG particle mass.
This function, quantified by the inverse slope 
parameter $C$ and the exponent parameter $n$, describes both the exponential shape of the spectrum at 
low \pT~and the power law distribution at large \pT. The parameter 
\dndy~represents the particle yield per unit rapidity per INEL event.
\dndy, $C$ and $n$ are the free parameters considered for this function.
Table~\ref{tab:levi} presents the parameter outcome of the \Tsallis~fit, 
together with the mean transverse momentum, \meanpT, and the reduced \chisquare.
\begin{table}[h!]
\centering
\small
\begin{tabular}{c|ccc|c|c}
\hline\noalign{\smallskip}
Baryon                & \dndy~(LT) {\scriptsize ($\times$10$^{-3}$)} & $C$ (MeV) & $n$  & \meanpT~(LT) {\scriptsize (\gmom)} & \reducedchisquare \\
\hline\noalign{\smallskip}
\rmSigma(1385)$^{+}$  & 9.8  $\pm$ 0.2 $\pm$ 0.9 & 301 $\pm$ 39  $\pm$ 15 & 9.0 $\pm$ 2.9 $\pm$ 0.5 &  1.17 $\pm$ 0.02 $\pm$ 0.03 & 1.13/7 \\
\rmSigma(1385)$^{-}$  & 10.6 $\pm$ 0.2 $\pm$ 1.1 & 308 $\pm$ 39  $\pm$ 20 & 9.1 $\pm$ 3.2 $\pm$ 0.8 &  1.17 $\pm$ 0.02 $\pm$ 0.03 & 1.71/7 \\
\rmASigma(1385)$^{-}$ & 9.0  $\pm$ 0.2 $\pm$ 0.9 & 307 $\pm$ 40  $\pm$ 15 & 9.8 $\pm$ 3.7 $\pm$ 0.8 &  1.18 $\pm$ 0.02 $\pm$ 0.04 & 1.19/7 \\
\rmASigma(1385)$^{+}$ & 10.0 $\pm$ 0.2 $\pm$ 1.1 & 294 $\pm$ 43  $\pm$ 17 & 8.9 $\pm$ 3.5 $\pm$ 0.6 &  1.18 $\pm$ 0.02 $\pm$ 0.04 & 1.53/7 \\
\hline\noalign{\smallskip}
\rmXiZres             & 2.48 $\pm$ 0.07 $\pm$ 0.24 & 404 $\pm$ 20 $\pm$ 21 & 16.9 $\pm$ 3.9 $\pm$ 1.9 &  1.33 $\pm$ 0.02 $\pm$ 0.05 & 2.24/5 \\
\noalign{\smallskip}\hline\noalign{\smallskip}
\end{tabular}
\caption{Parameters extracted from the \Tsallis~(LT) fits 
(Eq.~\ref{eqn:funclevy}) to the transverse momentum spectra.
The values of \dndy~are calculated using the spectra in the measured range and the extrapolation of the fitted
\Tsallis~function outside the measured range. Systematic uncertainties quoted here are the ones derived
from \Tsallis~fit only (see text).
}
\label{tab:levi}       %
\end{table}
The values of \dndy~in Table~\ref{tab:levi} are obtained by adding the integral of the experimental spectrum in the measured range 
and the extrapolations with the fitted \Tsallis~function to both \pT~$=0$ and high \pT.
The contribution of the low-\pT~extrapolation to the total \dndy~is $\sim$~30\% for both \rmSigma(1385) and \rmXiZres. 
The contribution of the high-\pT~extrapolation is negligible. 
For each species considered here, such a composite \dndy~differs very little ($<$ 1\%) from the value of \dndy~as the first
free parameter returned by the fit, i.e. from the integration of the fit function from 0 to infinity. 

In order to obtain the systematic uncertainty on the parameters of the \Tsallis~fit 
(\dndy, $C$ and $n$) and on the mean transverse momentum (\meanpT),
the \Tsallis~fit is repeated for each
\pT~spectrum obtained by varying separately the selection criteria in each
source of systematic uncertainties. Only statistical uncertainties
on the points of the \pT~spectrum are used for the fit. The values for
\dndy, $C$, $n$ and \meanpT, obtained for each source, are compared to
those from the fit to the reference \pT~spectrum, obtained
with default selection criteria. The fit to the reference 
\pT~spectrum is also done with statistical uncertainties only. 
The statistically significant differences are summed in quadrature
to contribute to the overall systematic uncertainties on \dndy, $C$, $n$ and \meanpT.

Although the \Tsallis~function describes the spectra both at low and at large \pT, 
other functions (e.g. $m_{\mathrm{T}}$ exponential or \pT~power law)
are likely to reproduce the low-\pT~behaviour and are suitable for the low-\pT~extrapolation.
These functions are fitted to the \mbox{low-\pT}~part of the spectrum below $3$~\gmom~and
used to evaluate the low-\pT~contribution outside the measured range.
An $m_{\mathrm{T}}$ exponential functional form
\begin{eqnarray}
  \frac{1}{N_{\rm INEL}}\frac{\mathrm{d}^{2}N}{\mathrm{d}y\mathrm{d}\pT}  &  =  &   A\; 
  \pT \;m_{\mathrm{T}} \;e^{-\frac{m_{\mathrm{T}}}{C}} \label{eqn:funcmtexp},
\end{eqnarray}
where $A$ is the normalization factor and $C$ is the inverse slope parameter,
gives values for \dndy~which are $\sim$~5-6\% lower and values for \meanpT~which
are $\sim$~3\% higher than those obtained with the \Tsallis~function.
A \pT~power law functional form %
\begin{eqnarray}
  \frac{1}{N_{\rm INEL}}\frac{\mathrm{d}^{2}N}{\mathrm{d}y\mathrm{d}\pT}  &  =  & A\; 
  \pT \; \left( 1+ \frac{\pT}{nC}\right)^{-n} \label{eqn:funcptpower},
\end{eqnarray}
gives values for \dndy~which are $\sim$~10-15\% higher and values for \meanpT~which
are $\sim$~9-11\% lower than those obtained with the \Tsallis~function.
Arithmetic averages of the values obtained with the three functions (\Tsallis,  $m_{\mathrm{T}}$ exponential, \pT~power law) 
are taken for \dndy~and \meanpT~and 
the unbiased estimators of standard deviation are considered as systematic uncertainties associated to the low-\pT~extrapolation.
These systematic uncertainties are summed in quadrature to contribute to the overall systematic 
uncertainties on \dndy~and \meanpT. Table \ref{tab:yields} summaries the results. 
\begin{table}[h!]
\centering
\begin{tabular}{c|c|c}
\hline\noalign{\smallskip}
Baryon                & \dndy~($\times$10$^{-3}$) & \meanpT~(\gmom) \\
\hline\noalign{\smallskip}
\rmSigma(1385)$^{+}$  & 10.0 $\pm$ 0.2 $^{+1.5} _{-1.4}$  & 1.15 $\pm$ 0.02 $\pm$ 0.07  \\
\rmSigma(1385)$^{-}$  & 10.8 $\pm$ 0.2 $^{+1.7} _{-1.6}$  & 1.15 $\pm$ 0.02 $\pm$ 0.08  \\
\rmASigma(1385)$^{-}$ & 9.1 $\pm$ 0.2 $^{+1.5} _{-1.4}$  &  1.16 $\pm$ 0.02 $\pm$ 0.08  \\
\rmASigma(1385)$^{+}$ & 10.3 $\pm$ 0.2 $^{+1.7} _{-1.5}$  & 1.16 $\pm$ 0.02 $\pm$ 0.07 \\
\hline\noalign{\smallskip}
\rmXiZres             & 2.56 $\pm$ 0.07 $^{+0.40} _{-0.37}$ & 1.31 $\pm$ 0.02 $\pm$ 0.09 \\
\noalign{\smallskip}\hline\noalign{\smallskip}
\end{tabular}
\caption{
Particle yield per unit rapidity, \dndy, and  mean transverse momentum, \meanpT. 
Values are obtained as an average of the values calculated with three different functions 
(\Tsallis~(Eq.~\ref{eqn:funclevy}),  $m_{\mathrm{T}}$ exponential (Eq.~\ref{eqn:funcmtexp}), 
\pT~power law (Eq.~\ref{eqn:funcptpower})), which reproduce the low-\pT~behaviour of the spectrum.
Systematic uncertainties include those from the low-\pT~extrapolation and (for \dndy~only) the 
\pT-independent uncertainties from Table \ref{tab:sys}.
}
\label{tab:yields}       %
\end{table}

The anti-baryon to baryon ratios, \rmASigma(1385)$^{-}/$\rmSigma(1385)$^{+}$ 
and \rmASigma(1385)$^{+}/$\rmSigma(1385)$^{-}$, 
are compatible with unity, although the large uncertainties leave very little predictive power on the
mechanisms of baryon-number transport~\cite{antibaryon-to-baryon}.  
%
\subsection{Comparison to models}
\label{sec:models}
\begin{figure}[t]
  \resizebox{1.0\columnwidth}{!}{
    \includegraphics{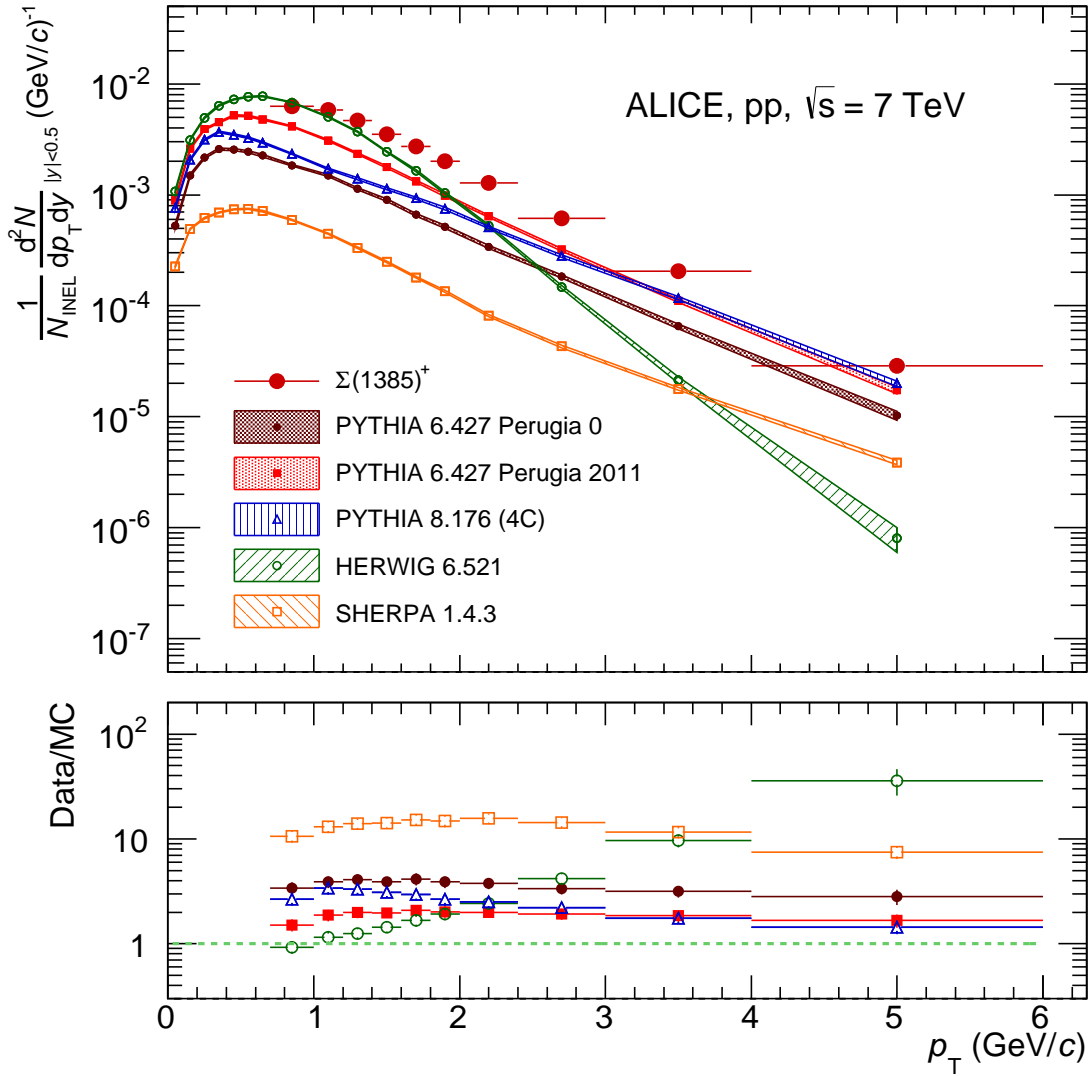}}
  \caption{The transverse momentum spectrum of \rmSigma(1385)$^{+}$ is 
    compared to 
    standard tunes of PYTHIA~6~\cite{PYTHIA} and PYTHIA~8~\cite{perugia8}, the latest release of HERWIG (6.521)~\cite{herwig}, and
    SHERPA release 1.4.6~\cite{sherpa}. 
    The MC data are binned according to the data.
    Spectra points are represented at the centre of the \pT~interval.
    The lower panel shows the ratio data/MC. \pT-independent uncertainties are not shown.
   }
  \label{fig:DataMCSigma}       %
\end{figure}

\begin{figure}[t]
  \resizebox{1.0\columnwidth}{!}{
    \includegraphics{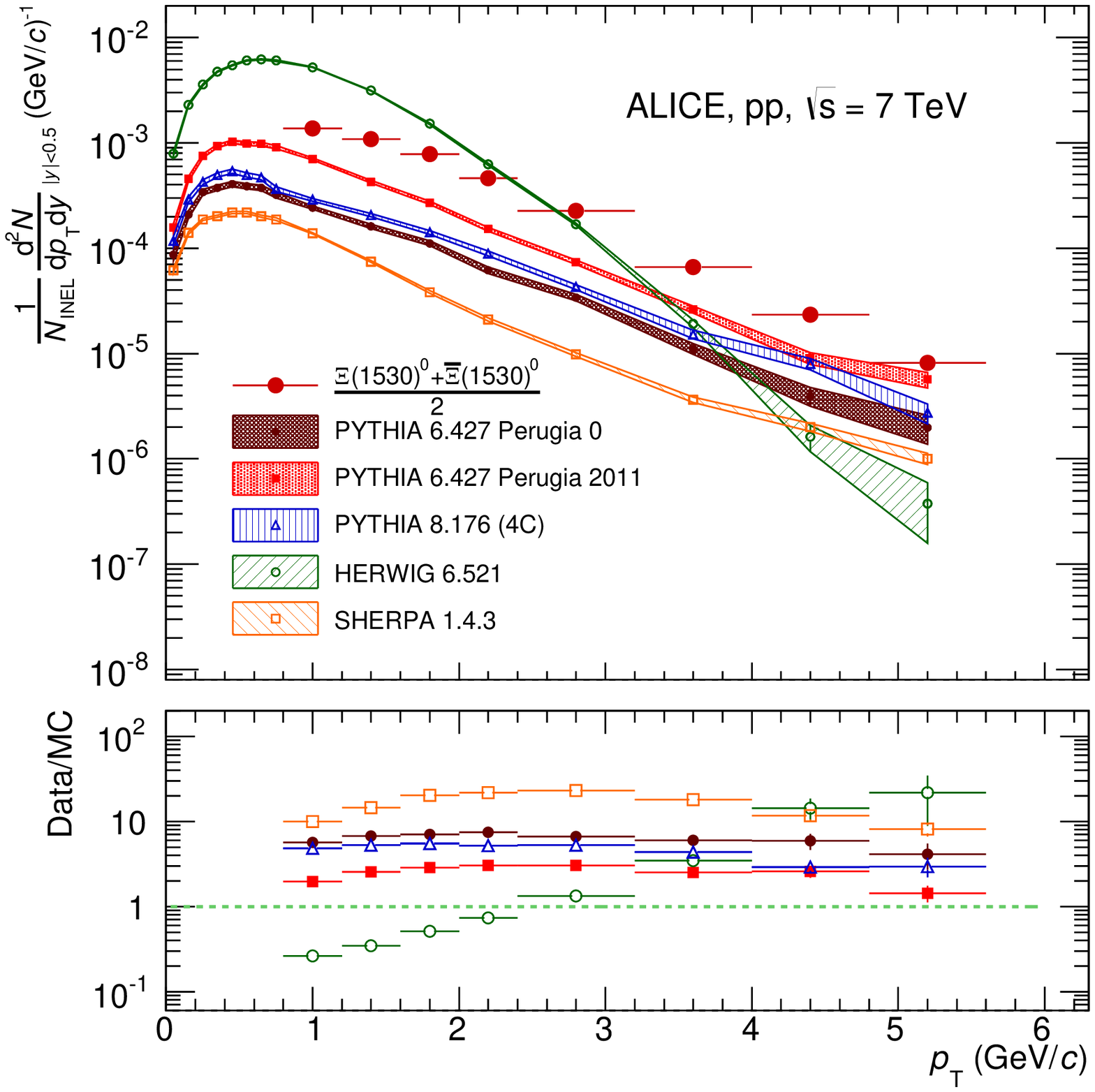}}
  \caption{
Same as Fig.~\ref{fig:DataMCSigma} but for  \rmXiZres.
}
  \label{fig:DataMCXi}       %
\end{figure}

The transverse momentum spectra of both \rmSigma(1385) and \rmXiZres~are 
compared to standard tunes of PYTHIA 6~\cite{PYTHIA} and PYTHIA~8~\cite{perugia8}, HERWIG~\cite{herwig} and SHERPA~\cite{sherpa}.
This is shown  in Figs.~\ref{fig:DataMCSigma} and~\ref{fig:DataMCXi} 
for \rmSigma(1385)$^{+}$ and \rmXiZres, respectively. 
Similar results to those of \rmSigma(1385)$^+$ are obtained for the other \rmSigma(1385) species. 

The latest release of PYTHIA~6 (6.427) is used. 
One of its latest tunes (Perugia~2011, tune~350~\cite{pythiatunes}) 
is compared with the central parameter set (Perugia~0, tune~320). 
Perugia~2011 takes into account some of the early LHC 
minimum-bias and underlying-event data at 0.9~TeV and 7~TeV 
(see~\cite{pythiatunes} and references therein) and describes
the 7~TeV \pp~charged particle spectra reasonably well~\cite{ALICE_charged}. 
The multi-strange baryon yields are also better described
by the Perugia~2011 tune, even if it still underpredicts the data~\cite{AliceMultiStrange}. 
Similar conclusions hold for the strange meson resonances $\phi$ and K$^*$~\cite{Abelev:2012}. 
For both \rmSigma(1385) and \rmXiZres, the Perugia~2011 tune underestimates the data,
though it gives a better description with respect to Perugia~0. 
Also the Perugia~2012 tune of PYTHIA~6 (tune~370~\cite{perugia2012}) has been tested 
with no significant improvement in the predictions for both \rmSigma(1385) and \rmXiZres.
The Perugia~2012 tune~\cite{perugia2012} is a retune of Perugia~2011 which
utilizes a different parton distribution function, CTEQ6L1 instead of CTEQ5L. 
The predictions from the Perugia~2012 tune are not reported in 
Figs.~\ref{fig:DataMCSigma} and~\ref{fig:DataMCXi}.

The latest release of PYTHIA~8 (8.176) is used. 
The standard 4C~tune (CTEQ6L1 \cite{perugia8}) gives a worse description 
with respect to the Perugia~2011 tune of PYTHIA~6.
The 4C~tune has color reconnection (CR) enabled by default: switching CR off
gives a worse description, as expected~\cite{color_reconnection}. 
ATLAS tunes A2-MSTW2008LO and AU2-CTEQ6L1 have been considered as 
alternatives to the standard 4C~tune (CTEQ6L1). The A2-MSTW2008LO utilizes a different
parton distribution function and the AU2-CTEQ6L1 is better tuned for underlying 
events. None of them performs better than the 4C~tune; therefore, they are not reported in 
Figs.~\ref{fig:DataMCSigma} and~\ref{fig:DataMCXi}.

Also shown in Figs.~\ref{fig:DataMCSigma} and~\ref{fig:DataMCXi} 
are the results from HERWIG (release 6.521)~\cite{herwig} and SHERPA (release 1.4.6)~\cite{sherpa}. 
HERWIG predicts a much softer production than for both the other models and the data,
for both \rmSigma(1385) and \rmXiZres.
For \rmSigma(1385), HERWIG is likely to describe the data at low \pT, but it
underpredicts the data by a factor $\sim$~2$-$4 in the intermediate-\pT~region $2<$~\pT~$<3$~\gmom,
and more than one order of magnitude at higher \pT.
For \rmXiZres, HERWIG fails both at low \pT, where the predictions are overestimated by a factor $\sim$~2$-$4,
and at high \pT, where the predictions are underestimated by more than one order of magnitude.
SHERPA gives a better description of the spectral shape
for both \rmSigma(1385) and \rmXiZres, but the overall
production cross sections are largely underestimated. 

The integrated yields \dndy~are also compared to thermal model calculations by
F.~Becattini $\it{et~al.}$~\cite{becattini}, 
tuned on the yields measured by the ALICE experiment at $\sqrt{s}$~=~7~TeV for 
$\pi^{+}$, K$^{*0}$, $\phi$, $\Xi^{\pm}$ and $\Omega^{\pm}$~\cite{ALICE_charged,Abelev:2012,AliceMultiStrange}, giving
a temperature of $T=$~160~MeV. 
The other parameters, as obtained from the fit to the ALICE data,
are the strangeness suppression factor, $\gamma_{S}$~=~0.72,
the normalization parameter, $A$~=~0.0355,  and $V T^{3}$~=~231.2, where $V$ is the volume.
The comparison is done for the ratios \rmSigma(1385)$^{+}$$/$\rmLambda~and \rmXiZres$/\Xi^{-}$,
which are sensitive to the temperature $T$. 
The experimental yields of \rmLambda~and $\Xi^{-}$ are from~\cite{AliceLambda7TeV,AliceMultiStrange}.
The theoretical value for the \rmXiZres~is obtained as average of the values 
for \rmXiZres~and \rmAXiZres, to be compared to the experimental results of this analysis.
The theoretical prediction for \rmSigma(1385)$^{+}$$/$\rmLambda~(0.13) is in agreement with the measured
value (0.131~$\pm$~0.002~$\pm$~0.021). Similar conclusions hold
for the other \rmSigma(1385) species (namely, for the ratios \rmSigma(1385)$^{-}$$/$\rmLambda,
\rmASigma(1385)$^{-}$$/$\rmALambda~and \rmASigma(1385)$^{+}$$/$\rmALambda).
The prediction for \rmXiZres$/\Xi^{-}$ (0.38) is also in agreement with
the experimental value (\mbox{$0.32 \pm 0.01 \pm 0.05$}) 
if both statistical and systematic uncertainties are considered.
%
\subsection{Mean transverse momentum \meanpT}
\label{meanpt}
\begin{figure}[!h]
\resizebox{1.\columnwidth}{!}{
\includegraphics{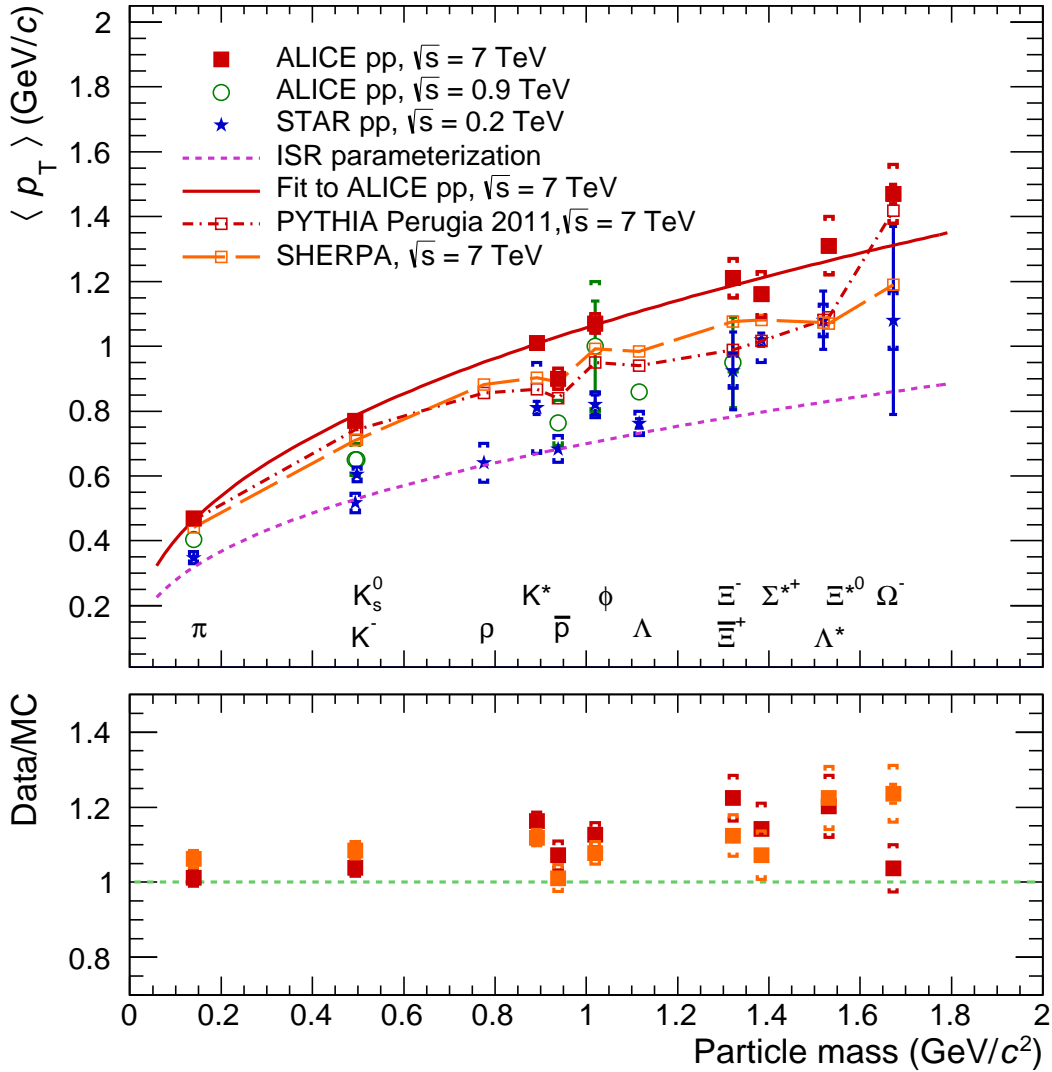}
}
\caption{The mean \pT~as function of the particle mass including \rmSigma(1385)$^+$, \rmXiZres~and 
other particles reconstructed in \pp~collisions 
at  \s~=~7 TeV and  \s~=~0.9~TeV by the ALICE collaboration~\cite{Aamodt:2011, Abelev:2012, Spectra_900,ALICE_charged,AliceMultiStrange}
and at  \s~=~0.2~TeV by the STAR collaboration~\cite{STAR_Identified,STAR_Phi,STAR_Strange,STAR_StrangeBaryons,STAR_2008}. 
The lower panel shows the ratio data/MC. Statistical and systematic uncertainties are shown separately (vertical
solid lines and brackets, respectively).
}
\label{fig:MeanPtVsMass}       %
\end{figure}

The mean transverse momentum \meanpT~serves as a single variable to
characterize the soft part of the measured particle spectra.
Figure~\ref{fig:MeanPtVsMass} shows the \meanpT~as a function of the particle mass,
covering a wide range of hadron mass up to the $\Omega^{-}$. 
The plot includes \rmSigma(1385)$^+$ and
\rmXiZres~from this analysis, and other particles measured in \pp~collisions at \s~=~0.9~TeV
and \s~=~7~TeV with the ALICE experiment~\cite{Aamodt:2011,Abelev:2012, Spectra_900,ALICE_charged,AliceMultiStrange}. 
The STAR \pp~data at \s~=~0.2~TeV~\cite{STAR_Identified,STAR_Phi,STAR_Strange,STAR_StrangeBaryons,STAR_2008} are added for comparison.
The dashed line in Fig.~\ref{fig:MeanPtVsMass} is the ISR parametrization,
an empirical curve proposed originally~\cite{ISRcurve} to
describe the ISR~\cite{ISR} and FNAL~\cite{FNAL} data for $\pi$, K and p only, at \s~=~0.025~TeV.

For STAR data, the ISR parametrization still works relatively well
for lower-mass particles up to \linebreak $\sim$~1~\Gmass~\cite{STAR_Strange}, 
despite the jump in the collision energy by
nearly an order of magnitude with respect to previous experiments, but it fails 
to describe the dependence of \meanpT~for 
higher-mass particles. At the RHIC energies, this was attributed 
to an increasing contribution to the transverse momentum spectra from mini-jet 
production~\cite{minijet}. In particular, it was noted that strange baryon resonances 
($\Sigma$(1385) and $\Lambda$(1520)) 
follow a steeper increase, similar to the trend of heavier mass particles~\cite{STAR_StrangeBaryons}.

For ALICE data, the ISR parametrization fails to fit the lower-mass particles already
at the collision energy of \s~=~0.9~TeV and the dependence of \meanpT~with the mass is even steeper at \s~=~7~TeV.
Unlike STAR, strange baryon resonances follow the same trend as the lower-mass particles. 
At the LHC energies, flow-like effects in pp collisions are investigated~\cite{epos, color_reconnection} 
which might explain the harder behaviour of
transverse momentum spectra, specially for higher mass particles. 
The ALICE points at \s~=~7~TeV are fitted with a function similar to the 
ISR parametrization, 
\begin{equation}
\langle \pT \rangle = \alpha \left(\frac{M}{1~\rm{GeV}/\it{c}^{\rm{2}}}\right)^{\beta}, 
\end{equation}
where $M$ is the particle mass,
obtaining $\alpha$~=~(1.06~$\pm$~0.02)~\gmom~and $\beta$~=~0.43~$\pm$~0.02. 
For the fit the statistical and systematic uncertainties are summed in quadrature. 
A \chisquare/ndf~=~9.61/6 with a probability of 14\%, is obtained. 
The antiproton \meanpT~is excluded from the fit since it is off-trend.
Including it in the fit changes very little the fit parameters 
($\alpha$~=~1.04~\gmom~and $\beta$~=~0.41) but increases the \chisquare~(\chisquare/ndf~=~15.75/7). 
The values for $\alpha$ and $\beta$ have to be compared
with $\alpha_{\rm ISR}$~=~0.7~\gmom~and $\beta_{\rm ISR}$~=~0.4. The results of the fit are shown with a 
solid line in Fig.~\ref{fig:MeanPtVsMass}. 

The dash-dotted line in Fig.~\ref{fig:MeanPtVsMass} is the prediction from PYTHIA~6, tune Perugia~2011. 
For \rmSigma(1385)$^+$ and \rmXiZres~the MC predictions
are $\sim$~20\% softer than data.
The long-dashed line in Fig.~\ref{fig:MeanPtVsMass} is the prediction from SHERPA, which is also
softer than data.
%
\section{Search for the $\phi(1860)$ pentaquark}
\label{pentaquark}

\begin{figure}[h!]
\resizebox{1.0\textwidth}{!}{
  \includegraphics{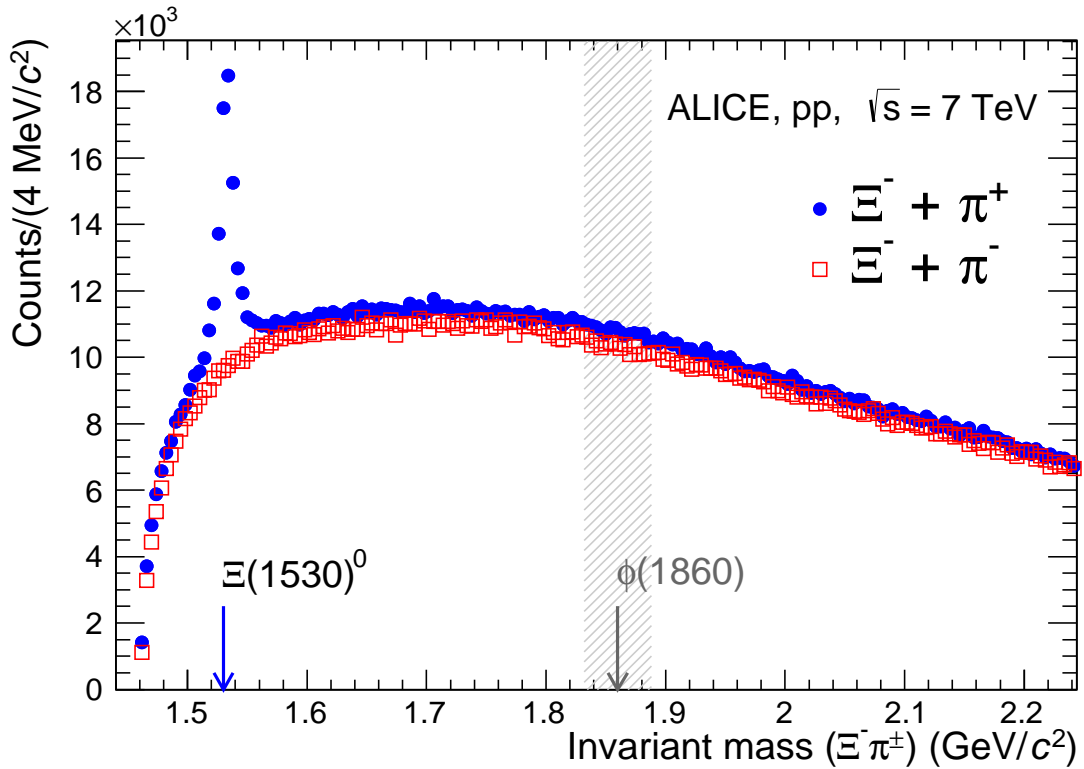}
}
\caption{\rmXi$^{-}$\piPlus~and \rmXi$^{-}$\piMinus~invariant mass distributions.
The arrow and the shaded area indicate the region where the $\phi(1860)$ pentaquark is 
expected and where the search was performed.}
\label{fig:PentaQ}      
\end{figure}

In order to explore the existence of the $\phi(1860)$ pentaquark, reported by the NA49 experiment~\cite{NA49}, 
the \rmXi$^{-}$\piPlus~invariant mass spectrum
in Fig.~\ref{fig:invmassXiStar} was extended up to above 2~\Gmass, as shown in Fig.~\ref{fig:PentaQ}.
The arrow and the shaded area give the region where the $\phi(1860)$ pentaquark is 
expected and where the search was performed.
From MC studies with reconstructed particles, the detector mass resolution of the \rmXiZres~is $\sim$~2~\mmass~and no
significant worsening is expected at masses around 1860~\mmass. The expected theoretical width of the $\phi(1860)$ is
quite narrow ($\lesssim$~10~\mmass~\cite{NA49}) so that, eventually, the detector resolution should not 
affect the measurement.
Also in Fig.~\ref{fig:PentaQ} the like-sign, \rmXi$^{-}$\piMinus, invariant mass distribution is presented. 
Both channels could potentially exhibit a signature of the  $\phi(1860)$ pentaquark: 
$\phi(1860)^{0}$ in the \rmXi$^{-}$\piPlus channel and
$\phi(1860)^{--}$ in the \rmXi$^{-}$\piMinus channel. Both distributions in Fig.~\ref{fig:PentaQ} 
clearly demonstrate the lack of significant evidence for the $\phi(1860)$ pentaquark.  

No signal of the pentaquark was found by many other 
experiments~\cite{ALEPH,BABAR,CDF,COMPASS,E690,FOCUS,HERA-B,HERMES,WA89,ZEUS,H1}.
A measure of the maximum likely yield of the $\phi(1860)$ has been made according to the procedure
used by the COMPASS experiment~\cite{COMPASS}.  The background is first estimated by fitting the like-sign distribution 
with a 4$^{\rm{th}}$-order polynomial from 1.6 to 2.2~\Gmass~while excluding the supposed pentaquark 
range from 1.825 to 1.895~\Gmass.  
The signal is counted by integrating the entries in the like-sign distribution in a 28~\mmass~interval 
centred around 1.860~\Gmass.
The maximum likely signal estimated at the 3~$\sigma$ ($99\%$) confidence level is
\begin{equation}
S_{\phi(1860)} = 3\sqrt{b} + \mathrm{max}(0,s-b),
\label{eq:PQeq}
\end{equation}
where the counted signal and background are given by $s$ and $b$, respectively.  
The ratio of the integrated \rmXi(1530)$^{0}$~yield to the pentaquark yield, $S_{\phi(1860)}$, is 
to be compared to other experiments. This is shown in Table~\ref{tab:penta} for the $\phi(1860)^{--}$.
The acceptance effects largely cancel in the ratio.
The pentaquark search was also performed moving the centre of the search interval by 
10~\mmass~to the left and to the right; the same result is obtained.
Similar results for $S_{\phi(1860)}$ are obtained for $\phi(1860)^{0}$.
\renewcommand{\thefootnote}{\textdagger}
\begin{savenotes}
\begin{table}[h!]
%
%
\centering
\begin{tabular}{l|c|c|r|r}
\hline\noalign{\smallskip}
Experiment & Initial state & Energy (TeV) & $S_{\phi(1860)^{--}}$ & \rmXiZres/$S_{\phi(1860)^{--}}$ \\
\hline\noalign{\smallskip}
ALICE                  & pp                & $\sqrt{s}$~=~7                & $<$807         & $>$44 \\
\hline\noalign{\smallskip}
NA49 \cite{NA49}       & pp                & $E_{\rm{p}}$~=~0.158           & 36        & 4.2  \\
\hline\noalign{\smallskip}
ALEPH \cite{ALEPH}     &  e$^+$e$^-$       & $\sqrt{s}$~=~$m_{\rm{Z}^{0}}$   & $<$24     & $>$13.4 \\
BaBar \cite{BABAR}     &  e$^+$e$^-$       & $\sqrt{s}$~=~$m_{\Upsilon(4S)}$ & not seen  &  \\
CDF \cite{CDF}         &  p$\bar{\rm{p}}$  & $\sqrt{s}$~=~1.960            & $<$63     &  $>$35 \\
COMPASS \cite{COMPASS} &  $\mu^+$--A       & $E_{\rm{\mu^{+}}}$~=~0.160       & $<$79     & $>$ 21.5 \\
E690 \cite{E690}       &  pp               & $E_{\rm{p}}$~=~0.800           & $<$310    & $>$302  \\
FOCUS \cite{FOCUS}     &  $\gamma$p        & $E_{\gamma} \le$~0.300       & $<$170    & $>$349 \\
HERA-B \cite{HERA-B}   &  p--A             & $E_{\rm{p}}$~=~0.920           & $<$56     & $>$25 \\
HERMES \cite{HERMES}   &  e$^-$--D         & $E_{\rm{e}}$~=~0.0276          &$<$5       & $>$7 \\
WA89 \cite{WA89}       &  $\Sigma^-$--A    & $E_{\rm{\Sigma^{-}}}$~=~0.340    & $<$760    & $>$79 \\
ZEUS \cite{ZEUS}       &  ep               & $\sqrt{s}$~=~0.300, 0.318            & not seen  &  \\
H1 \cite{H1}       &  ep               & $\sqrt{s}$~=~0.300, 0.318            & not seen  & $>$2-8\footnote[1]{At the 95\% C.L.} \\
\hline\noalign{\smallskip}
\noalign{\smallskip}
\end{tabular}
\caption{Summary of $\phi$(1860) searches in inclusive production. The energies given in the third column 
refer to the beam energy in case of fixed-target experiments and to $\sqrt{s}$ in case of 
collider experiments. The pentaquark signal is related to the \rmXiZres~yield in the last column.
}
\label{tab:penta}
%
%
\end{table}
\end{savenotes}
%
\section{Conclusions}
\label{conclusions}
The transverse momentum spectra of the baryon resonances \rmSigma(1385) and \rmXiZres~have been  
measured by the ALICE collaboration in pp collisions at an energy in the centre of mass of \s~=~7~TeV. 
A \Tsallis~function describes the spectra well.

The mean transverse momentum \meanpT~of both \rmSigma(1385) and \rmXiZres,
when plotted as a function of the particle mass, follows
the trend of other particles measured with the ALICE experiment
in \pp~collisions at \s~=~7~TeV. 

The differential spectra have been compared to several MC event generators, e.g.
standard tunes of PYTHIA~6 and PYTHIA~8, HERWIG and
SHERPA. PYTHIA~6 Perugia~2011
(tune~350) performs better than any other tested generator, still underpredicting
the data by a factor $\sim$~2-3 in the intermediate-\pT~region $2<$~\pT~$<3$~\gmom.

The search for the $\phi(1860)^{0}$ and $\phi(1860)^{--}$ pentaquark states in the $\Xi\pi$ charged channels 
has shown no evidence for the existence of such exotic particles.

%
%
\newenvironment{acknowledgement}{\relax}{\relax}
\begin{acknowledgement}
\section*{Acknowledgements}
\input{acknowledgements_march2013.tex}    
\end{acknowledgement}
%
%
%
\bibliography{mybib}{}
\bibliographystyle{utphys}
\newpage
%
\appendix
\section{The ALICE Collaboration}
\label{app:collab}
\input{Alice_Authorlist_2014-Jun-03-CERNPREP.tex}  
\end{document}

%% file: acknowledgements_march2013.tex
The ALICE collaboration would like to thank all its engineers and technicians for their invaluable contributions to the construction of the experiment and the CERN accelerator teams for the outstanding performance of the LHC complex.
The ALICE collaboration acknowledges the following funding agencies for their support in building and
running the ALICE detector:
State Committee of Science,  World Federation of Scientists (WFS)
and Swiss Fonds Kidagan, Armenia,
Conselho Nacional de Desenvolvimento Cient\'{\i}fico e Tecnol\'{o}gico (CNPq), Financiadora de Estudos e Projetos (FINEP),
Funda\c{c}\~{a}o de Amparo \`{a} Pesquisa do Estado de S\~{a}o Paulo (FAPESP);
National Natural Science Foundation of China (NSFC), the Chinese Ministry of Education (CMOE)
and the Ministry of Science and Technology of China (MSTC);
Ministry of Education and Youth of the Czech Republic;
Danish Natural Science Research Council, the Carlsberg Foundation and the Danish National Research Foundation;
The European Research Council under the European Community's Seventh Framework Programme;
Helsinki Institute of Physics and the Academy of Finland;
French CNRS-IN2P3, the `Region Pays de Loire', `Region Alsace', `Region Auvergne' and CEA, France;
German BMBF and the Helmholtz Association;
General Secretariat for Research and Technology, Ministry of
Development, Greece;
Hungarian OTKA and National Office for Research and Technology (NKTH);
Department of Atomic Energy and Department of Science and Technology of the Government of India;
Istituto Nazionale di Fisica Nucleare (INFN) and Centro Fermi -
Museo Storico della Fisica e Centro Studi e Ricerche "Enrico
Fermi", Italy;
MEXT Grant-in-Aid for Specially Promoted Research, Ja\-pan;
Joint Institute for Nuclear Research, Dubna;
National Research Foundation of Korea (NRF);
CONACYT, DGAPA, M\'{e}xico, ALFA-EC and the EPLANET Program
(European Particle Physics Latin American Network)
Stichting voor Fundamenteel Onderzoek der Materie (FOM) and the Nederlandse Organisatie voor Wetenschappelijk Onderzoek (NWO), Netherlands;
Research Council of Norway (NFR);
Polish Ministry of Science and Higher Education;
National Authority for Scientific Research - NASR (Autoritatea Na\c{t}ional\u{a} pentru Cercetare \c{S}tiin\c{t}ific\u{a} - ANCS);
Ministry of Education and Science of Russian Federation, Russian
Academy of Sciences, Russian Federal Agency of Atomic Energy,
Russian Federal Agency for Science and Innovations and The Russian
Foundation for Basic Research;
Ministry of Education of Slovakia;
Department of Science and Technology, South Africa;
CIEMAT, EELA, Ministerio de Econom\'{i}a y Competitividad (MINECO) of Spain, Xunta de Galicia (Conseller\'{\i}a de Educaci\'{o}n),
CEA\-DEN, Cubaenerg\'{\i}a, Cuba, and IAEA (International Atomic Energy Agency);
Swedish Research Council (VR) and Knut $\&$ Alice Wallenberg
Foundation (KAW);
Ukraine Ministry of Education and Science;
United Kingdom Science and Technology Facilities Council (STFC);
The United States Department of Energy, the United States National
Science Foundation, the State of Texas, and the State of Ohio.

%% file: Alice_Authorlist_2014-Jun-03-CERNPREP.tex


\begingroup
\small
\begin{flushleft}
B.~Abelev\Irefn{org71}\And
J.~Adam\Irefn{org37}\And
D.~Adamov\'{a}\Irefn{org79}\And
M.M.~Aggarwal\Irefn{org83}\And
G.~Aglieri~Rinella\Irefn{org34}\And
M.~Agnello\Irefn{org107}\textsuperscript{,}\Irefn{org90}\And
A.~Agostinelli\Irefn{org26}\And
N.~Agrawal\Irefn{org44}\And
Z.~Ahammed\Irefn{org126}\And
N.~Ahmad\Irefn{org18}
% ...
\renewcommand\labelenumi{\textsuperscript{\theenumi}~}

\section*{Affiliation notes}
\renewcommand\theenumi{\roman{enumi}}
\begin{Authlist}
\item \Adef{0}Deceased
\item \Adef{idp1133712}{Also at: St. Petersburg State Polytechnical University}
\item \Adef{idp3056368}{Also at: Department of Applied Physics, Aligarh Muslim University, Aligarh, India}
\item \Adef{idp3736992}{Also at: M.V. Lomonosov Moscow State University, D.V. Skobeltsyn Institute of Nuclear Physics, Moscow, Russia}
\item \Adef{idp3987568}{Also at: University of Belgrade, Faculty of Physics and "Vin\v{c}a" Institute of Nuclear Sciences, Belgrade, Serbia}
\item \Adef{idp4306272}{Permanent Address: Permanent Address: Konkuk University, Seoul, Korea}
\item \Adef{idp4870032}{Also at: Institute of Theoretical Physics, University of Wroclaw, Wroclaw, Poland}
\item \Adef{idp5796768}{Also at: University of Kansas, Lawrence, KS, United States}
\end{Authlist}

\section*{Collaboration Institutes}
\renewcommand\theenumi{\arabic{enumi}~}
\begin{Authlist}

\item \Idef{org1}A.I. Alikhanyan National Science Laboratory (Yerevan Physics Institute) Foundation, Yerevan, Armenia
\item \Idef{org2}Benem\'{e}rita Universidad Aut\'{o}noma de Puebla, Puebla, Mexico
\item \Idef{org3}Bogolyubov Institute for Theoretical Physics, Kiev, Ukraine
\item \Idef{org4}Bose Institute, Department of Physics and Centre for Astroparticle Physics and Space Science (CAPSS), Kolkata, India
\item \Idef{org5}Budker Institute for Nuclear Physics, Novosibirsk, Russia
\item \Idef{org6}California Polytechnic State University, San Luis Obispo, CA, United States
\item \Idef{org7}Central China Normal University, Wuhan, China
\item \Idef{org8}Centre de Calcul de l'IN2P3, Villeurbanne, France
\item \Idef{org9}Centro de Aplicaciones Tecnol\'{o}gicas y Desarrollo Nuclear (CEADEN), Havana, Cuba
\item \Idef{org10}Centro de Investigaciones Energ\'{e}ticas Medioambientales y Tecnol\'{o}gicas (CIEMAT), Madrid, Spain
\item \Idef{org11}Centro de Investigaci\'{o}n y de Estudios Avanzados (CINVESTAV), Mexico City and M\'{e}rida, Mexico
\item \Idef{org12}Centro Fermi - Museo Storico della Fisica e Centro Studi e Ricerche ``Enrico Fermi'', Rome, Italy
\item \Idef{org13}Chicago State University, Chicago, USA
\item \Idef{org14}Commissariat \`{a} l'Energie Atomique, IRFU, Saclay, France
\item \Idef{org15}COMSATS Institute of Information Technology (CIIT), Islamabad, Pakistan
\item \Idef{org16}Departamento de F\'{\i}sica de Part\'{\i}culas and IGFAE, Universidad de Santiago de Compostela, Santiago de Compostela, Spain
\item \Idef{org17}Department of Physics and Technology, University of Bergen, Bergen, Norway
\item \Idef{org18}Department of Physics, Aligarh Muslim University, Aligarh, India
\item \Idef{org19}Department of Physics, Ohio State University, Columbus, OH, United States
\item \Idef{org20}Department of Physics, Sejong University, Seoul, South Korea
\item \Idef{org21}Department of Physics, University of Oslo, Oslo, Norway
\item \Idef{org22}Dipartimento di Fisica dell'Universit\`{a} 'La Sapienza' and Sezione INFN Rome, Italy
\item \Idef{org23}Dipartimento di Fisica dell'Universit\`{a} and Sezione INFN, Cagliari, Italy
\item \Idef{org24}Dipartimento di Fisica dell'Universit\`{a} and Sezione INFN, Trieste, Italy
\item \Idef{org25}Dipartimento di Fisica dell'Universit\`{a} and Sezione INFN, Turin, Italy
\item \Idef{org26}Dipartimento di Fisica e Astronomia dell'Universit\`{a} and Sezione INFN, Bologna, Italy
\item \Idef{org27}Dipartimento di Fisica e Astronomia dell'Universit\`{a} and Sezione INFN, Catania, Italy
\item \Idef{org28}Dipartimento di Fisica e Astronomia dell'Universit\`{a} and Sezione INFN, Padova, Italy
\item \Idef{org29}Dipartimento di Fisica `E.R.~Caianiello' dell'Universit\`{a} and Gruppo Collegato INFN, Salerno, Italy
\item \Idef{org30}Dipartimento di Scienze e Innovazione Tecnologica dell'Universit\`{a} del  Piemonte Orientale and Gruppo Collegato INFN, Alessandria, Italy
\item \Idef{org31}Dipartimento Interateneo di Fisica `M.~Merlin' and Sezione INFN, Bari, Italy
\item \Idef{org32}Division of Experimental High Energy Physics, University of Lund, Lund, Sweden
\item \Idef{org33}Eberhard Karls Universit\"{a}t T\"{u}bingen, T\"{u}bingen, Germany
\item \Idef{org34}European Organization for Nuclear Research (CERN), Geneva, Switzerland
\item \Idef{org35}Faculty of Engineering, Bergen University College, Bergen, Norway
\item \Idef{org36}Faculty of Mathematics, Physics and Informatics, Comenius University, Bratislava, Slovakia
\item \Idef{org37}Faculty of Nuclear Sciences and Physical Engineering, Czech Technical University in Prague, Prague, Czech Republic
\item \Idef{org38}Faculty of Science, P.J.~\v{S}af\'{a}rik University, Ko\v{s}ice, Slovakia
\item \Idef{org39}Frankfurt Institute for Advanced Studies, Johann Wolfgang Goethe-Universit\"{a}t Frankfurt, Frankfurt, Germany
\item \Idef{org40}Gangneung-Wonju National University, Gangneung, South Korea
\item \Idef{org41}Gauhati University, Department of Physics, Guwahati, India
\item \Idef{org42}Helsinki Institute of Physics (HIP), Helsinki, Finland
\item \Idef{org43}Hiroshima University, Hiroshima, Japan
\item \Idef{org44}Indian Institute of Technology Bombay (IIT), Mumbai, India
\item \Idef{org45}Indian Institute of Technology Indore, Indore (IITI), India
\item \Idef{org46}Inha University, Incheon, South Korea
\item \Idef{org47}Institut de Physique Nucl\'eaire d'Orsay (IPNO), Universit\'e Paris-Sud, CNRS-IN2P3, Orsay, France
\item \Idef{org48}Institut f\"{u}r Informatik, Johann Wolfgang Goethe-Universit\"{a}t Frankfurt, Frankfurt, Germany
\item \Idef{org49}Institut f\"{u}r Kernphysik, Johann Wolfgang Goethe-Universit\"{a}t Frankfurt, Frankfurt, Germany
\item \Idef{org50}Institut f\"{u}r Kernphysik, Westf\"{a}lische Wilhelms-Universit\"{a}t M\"{u}nster, M\"{u}nster, Germany
\item \Idef{org51}Institut Pluridisciplinaire Hubert Curien (IPHC), Universit\'{e} de Strasbourg, CNRS-IN2P3, Strasbourg, France
\item \Idef{org52}Institute for Nuclear Research, Academy of Sciences, Moscow, Russia
\item \Idef{org53}Institute for Subatomic Physics of Utrecht University, Utrecht, Netherlands
\item \Idef{org54}Institute for Theoretical and Experimental Physics, Moscow, Russia
\item \Idef{org55}Institute of Experimental Physics, Slovak Academy of Sciences, Ko\v{s}ice, Slovakia
\item \Idef{org56}Institute of Physics, Academy of Sciences of the Czech Republic, Prague, Czech Republic
\item \Idef{org57}Institute of Physics, Bhubaneswar, India
\item \Idef{org58}Institute of Space Science (ISS), Bucharest, Romania
\item \Idef{org59}Instituto de Ciencias Nucleares, Universidad Nacional Aut\'{o}noma de M\'{e}xico, Mexico City, Mexico
\item \Idef{org60}Instituto de F\'{\i}sica, Universidad Nacional Aut\'{o}noma de M\'{e}xico, Mexico City, Mexico
\item \Idef{org61}iThemba LABS, National Research Foundation, Somerset West, South Africa
\item \Idef{org62}Joint Institute for Nuclear Research (JINR), Dubna, Russia
\item \Idef{org63}Konkuk University, Seoul, South Korea
\item \Idef{org64}Korea Institute of Science and Technology Information, Daejeon, South Korea
\item \Idef{org65}KTO Karatay University, Konya, Turkey
\item \Idef{org66}Laboratoire de Physique Corpusculaire (LPC), Clermont Universit\'{e}, Universit\'{e} Blaise Pascal, CNRS--IN2P3, Clermont-Ferrand, France
\item \Idef{org67}Laboratoire de Physique Subatomique et de Cosmologie, Universit\'{e} Grenoble-Alpes, CNRS-IN2P3, Grenoble, France
\item \Idef{org68}Laboratori Nazionali di Frascati, INFN, Frascati, Italy
\item \Idef{org69}Laboratori Nazionali di Legnaro, INFN, Legnaro, Italy
\item \Idef{org70}Lawrence Berkeley National Laboratory, Berkeley, CA, United States
\item \Idef{org71}Lawrence Livermore National Laboratory, Livermore, CA, United States
\item \Idef{org72}Moscow Engineering Physics Institute, Moscow, Russia
\item \Idef{org73}National Centre for Nuclear Studies, Warsaw, Poland
\item \Idef{org74}National Institute for Physics and Nuclear Engineering, Bucharest, Romania
\item \Idef{org75}National Institute of Science Education and Research, Bhubaneswar, India
\item \Idef{org76}Niels Bohr Institute, University of Copenhagen, Copenhagen, Denmark
\item \Idef{org77}Nikhef, National Institute for Subatomic Physics, Amsterdam, Netherlands
\item \Idef{org78}Nuclear Physics Group, STFC Daresbury Laboratory, Daresbury, United Kingdom
\item \Idef{org79}Nuclear Physics Institute, Academy of Sciences of the Czech Republic, \v{R}e\v{z} u Prahy, Czech Republic
\item \Idef{org80}Oak Ridge National Laboratory, Oak Ridge, TN, United States
\item \Idef{org81}Petersburg Nuclear Physics Institute, Gatchina, Russia
\item \Idef{org82}Physics Department, Creighton University, Omaha, NE, United States
\item \Idef{org83}Physics Department, Panjab University, Chandigarh, India
\item \Idef{org84}Physics Department, University of Athens, Athens, Greece
\item \Idef{org85}Physics Department, University of Cape Town, Cape Town, South Africa
\item \Idef{org86}Physics Department, University of Jammu, Jammu, India
\item \Idef{org87}Physics Department, University of Rajasthan, Jaipur, India
\item \Idef{org88}Physik Department, Technische Universit\"{a}t M\"{u}nchen, Munich, Germany
\item \Idef{org89}Physikalisches Institut, Ruprecht-Karls-Universit\"{a}t Heidelberg, Heidelberg, Germany
\item \Idef{org90}Politecnico di Torino, Turin, Italy
\item \Idef{org91}Purdue University, West Lafayette, IN, United States
\item \Idef{org92}Pusan National University, Pusan, South Korea
\item \Idef{org93}Research Division and ExtreMe Matter Institute EMMI, GSI Helmholtzzentrum f\"ur Schwerionenforschung, Darmstadt, Germany
\item \Idef{org94}Rudjer Bo\v{s}kovi\'{c} Institute, Zagreb, Croatia
\item \Idef{org95}Russian Federal Nuclear Center (VNIIEF), Sarov, Russia
\item \Idef{org96}Russian Research Centre Kurchatov Institute, Moscow, Russia
\item \Idef{org97}Saha Institute of Nuclear Physics, Kolkata, India
\item \Idef{org98}School of Physics and Astronomy, University of Birmingham, Birmingham, United Kingdom
\item \Idef{org99}Secci\'{o}n F\'{\i}sica, Departamento de Ciencias, Pontificia Universidad Cat\'{o}lica del Per\'{u}, Lima, Peru
\item \Idef{org100}Sezione INFN, Bari, Italy
\item \Idef{org101}Sezione INFN, Bologna, Italy
\item \Idef{org102}Sezione INFN, Cagliari, Italy
\item \Idef{org103}Sezione INFN, Catania, Italy
\item \Idef{org104}Sezione INFN, Padova, Italy
\item \Idef{org105}Sezione INFN, Rome, Italy
\item \Idef{org106}Sezione INFN, Trieste, Italy
\item \Idef{org107}Sezione INFN, Turin, Italy
\item \Idef{org108}SSC IHEP of NRC Kurchatov institute, Protvino, Russia
\item \Idef{org109}SUBATECH, Ecole des Mines de Nantes, Universit\'{e} de Nantes, CNRS-IN2P3, Nantes, France
\item \Idef{org110}Suranaree University of Technology, Nakhon Ratchasima, Thailand
\item \Idef{org111}Technical University of Split FESB, Split, Croatia
\item \Idef{org112}The Henryk Niewodniczanski Institute of Nuclear Physics, Polish Academy of Sciences, Cracow, Poland
\item \Idef{org113}The University of Texas at Austin, Physics Department, Austin, TX, USA
\item \Idef{org114}Universidad Aut\'{o}noma de Sinaloa, Culiac\'{a}n, Mexico
\item \Idef{org115}Universidade de S\~{a}o Paulo (USP), S\~{a}o Paulo, Brazil
\item \Idef{org116}Universidade Estadual de Campinas (UNICAMP), Campinas, Brazil
\item \Idef{org117}University of Houston, Houston, TX, United States
\item \Idef{org118}University of Jyv\"{a}skyl\"{a}, Jyv\"{a}skyl\"{a}, Finland
\item \Idef{org119}University of Liverpool, Liverpool, United Kingdom
\item \Idef{org120}University of Tennessee, Knoxville, TN, United States
\item \Idef{org121}University of Tokyo, Tokyo, Japan
\item \Idef{org122}University of Tsukuba, Tsukuba, Japan
\item \Idef{org123}University of Zagreb, Zagreb, Croatia
\item \Idef{org124}Universit\'{e} de Lyon, Universit\'{e} Lyon 1, CNRS/IN2P3, IPN-Lyon, Villeurbanne, France
\item \Idef{org125}V.~Fock Institute for Physics, St. Petersburg State University, St. Petersburg, Russia
\item \Idef{org126}Variable Energy Cyclotron Centre, Kolkata, India
\item \Idef{org127}Vestfold University College, Tonsberg, Norway
\item \Idef{org128}Warsaw University of Technology, Warsaw, Poland
\item \Idef{org129}Wayne State University, Detroit, MI, United States
\item \Idef{org130}Wigner Research Centre for Physics, Hungarian Academy of Sciences, Budapest, Hungary
\item \Idef{org131}Yale University, New Haven, CT, United States
\item \Idef{org132}Yonsei University, Seoul, South Korea
\item \Idef{org133}Zentrum f\"{u}r Technologietransfer und Telekommunikation (ZTT), Fachhochschule Worms, Worms, Germany
\end{Authlist}
\endgroup